\documentclass{article}
\usepackage{gen2vc_preprint,times}

\usepackage{amsmath,amsfonts,bm}

\def\eqref#1{equation~\ref{#1}}

\def\1{\bm{1}}

\DeclareMathAlphabet{\mathsfit}{\encodingdefault}{\sfdefault}{m}{sl}
\SetMathAlphabet{\mathsfit}{bold}{\encodingdefault}{\sfdefault}{bx}{n}

 \par

 \par

\usepackage{graphicx}
\usepackage{capt-of}
\usepackage{booktabs}
\usepackage{colortbl}
\usepackage{placeins}
\definecolor{tablegray}{gray}{0.92}
\makeatletter
\let\genvc@mainaddtocurcol\@addtocurcol
\makeatother
\usepackage{flafter}
\makeatletter
\let\genvc@appendixaddtocurcol\@addtocurcol
\let\@addtocurcol\genvc@mainaddtocurcol
\newcommand{\appendixfloatplacement}{\let\@addtocurcol\genvc@appendixaddtocurcol}
\makeatother
\usepackage{hyperref}
\usepackage{url}
\hypersetup{hidelinks}

\makeatletter
\def\fps@figure{!t}
\makeatother

\title{Gen2-VC: Unlocking Generative Priors\\for Video Compression}
\author{Yinhuan Huang, Jingkai Ying, Pu Chen, Zhijin Qin\thanks{Corresponding author.}\\
\normalfont Tsinghua University\\
\normalfont\small\texttt{huangyh24@mails.tsinghua.edu.cn}\quad\texttt{qinzhijin@tsinghua.edu.cn}}

\begin{document}
\maketitle
\noindent\begin{minipage}{\linewidth}
  \centering
  \includegraphics[width=\linewidth]{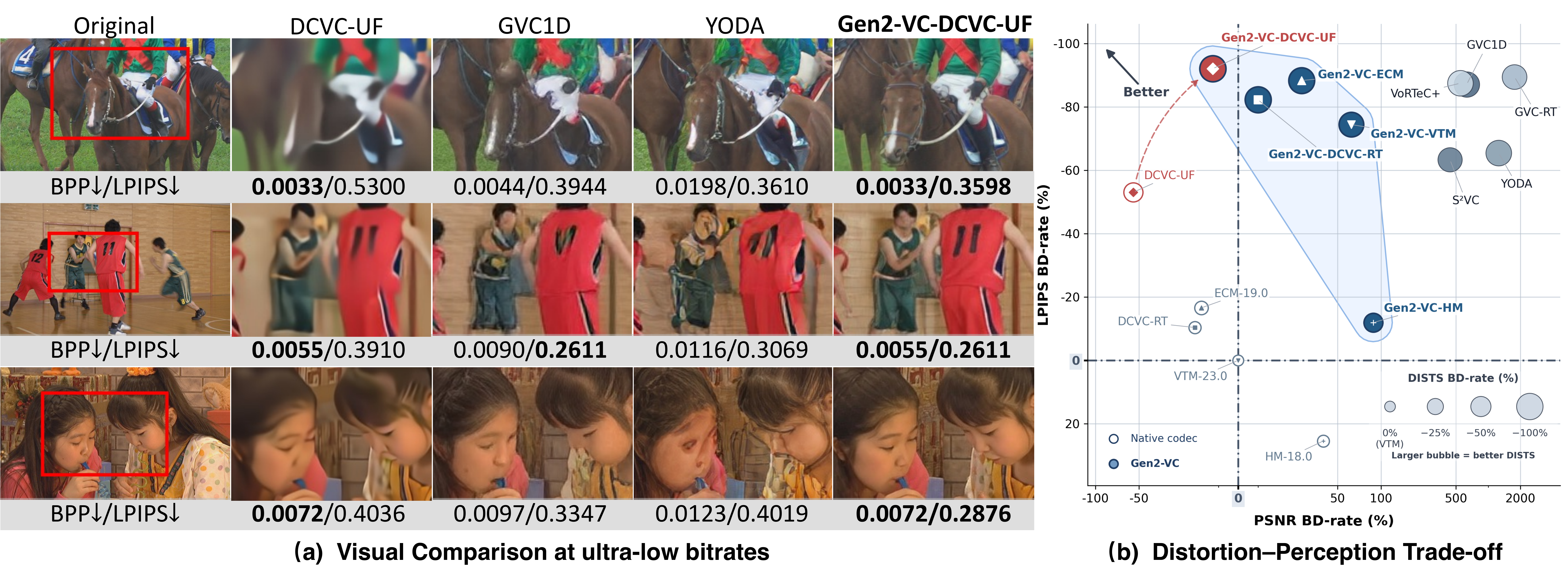}
  \captionof{figure}{(a) Gen2-VC better preserves source structure and details at the same or lower bitrates than distortion-oriented DCVC-UF~\citep{dcvcuf2026}, tokenizer-based GVC1D~\citep{gvc1d2026}, and diffusion-based YODA~\citep{yoda2026}, as illustrated by HEVC-D examples.
(b) Gen2-VC improves distortion--perception trade-offs across codecs through training-free transfer of adapters trained only on DCVC-UF, as illustrated on HEVC-E.
BD-rates (\%) $\downarrow$ are relative to VTM-23.0.}
  \label{fig:motivation}
\end{minipage}
\par\medskip
\begin{abstract}
Under stringent bitrate constraints, existing video codecs struggle to balance source fidelity and perceptual realism. 
Distortion-oriented codecs often oversmooth details, while generative codecs risk introducing content and structural deviations and rely on codec-specific designs. 
This motivates a question:
\emph{Can existing codecs achieve a better distortion--perception trade-off through simple, reusable adaptation?}
Our insight is that native codec reconstructions provide a shared interface through which generative refinement is anchored to source content while remaining decoupled from codec-specific representations.
We therefore propose Gen2-VC, a generative video compression framework that enhances the outputs of learned and conventional codecs with a pretrained video prior, leaving their bitstreams and reference update processes unchanged. 
With the codec, VAE, and generative backbone frozen, lightweight LoRA adapters refine codec reconstruction through single-frame spatial adaptation followed by multi-frame temporal adaptation with the video prior.
Using Wan2.1-T2V-1.3B, Gen2-VC-DCVC-UF outperforms previous leading codecs in LPIPS/DISTS and, to our knowledge, is the first generative video codec to surpass VTM-23.0 in both PSNR and MS-SSIM at low bitrates, based on BD-rates averaged over six datasets. 
Compared to VTM-23.0, it reduces bitrate by an average of 86.65\% and 94.24\% at matched LPIPS and DISTS, respectively. 
Adapters trained only on DCVC-UF improve perceptual quality on DCVC-RT, ECM, VTM, and HM without retraining.
\end{abstract}
\section{Introduction}
\label{sec:introduction}

Learned video compression has achieved impressive coding performance through deep models that capture spatial structures and temporal dependencies in compact representations.
However, stringent bitrate constraints limit how much source information these representations can retain.
Distortion-oriented codecs~\citep{dvc2019,fvc2021,dcvcuf2026} prioritize pixel-level reconstruction accuracy, but can oversmooth textures and fine details, as illustrated in Figure~\ref{fig:motivation}(a).
Even when coarse structures are preserved, missing details can leave reconstructed regions looking unnatural.
This motivates codec optimization that accounts for perceptual realism alongside pixel-level fidelity.

\begin{figure}[!t]
  \centering
  \includegraphics[width=\linewidth]{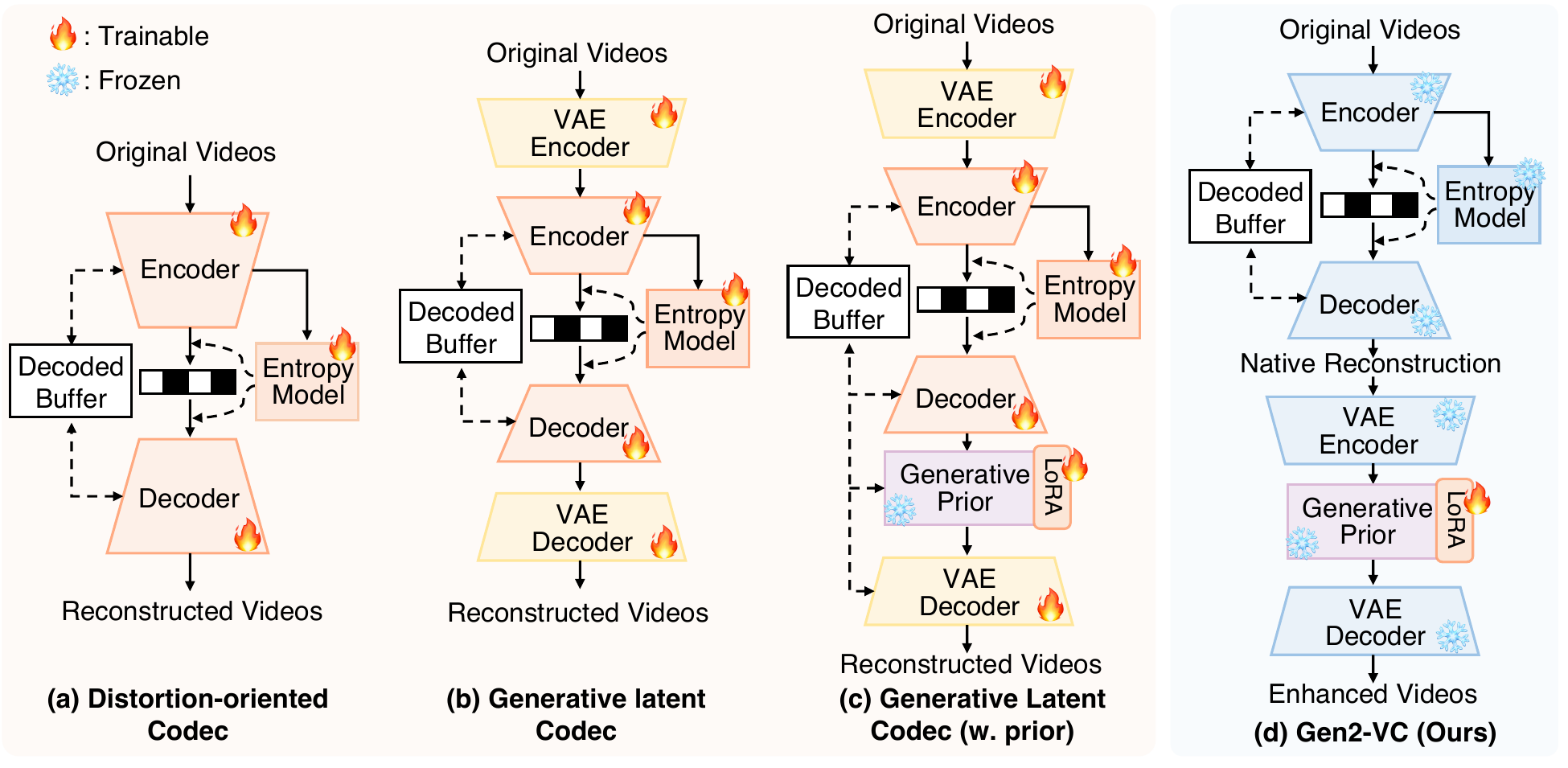}
  \caption{Comparison of video codec architectures.
  (a) Distortion-oriented coding reconstructs frames directly.
  (b) Generative latent coding uses perceptually or adversarially trained decoders.
  (c) Diffusion models refine decoded latents within conditional reconstruction pipelines.
  (d) Gen2-VC uses native reconstructions as a shared pixel-space interface, requiring no additional source-specific signals and preserving the codec's bitstream and reference updates.}
  \label{fig:interfaces}
\end{figure}

Generative video codecs seek to mitigate oversmoothing and improve texture realism.
One line compresses videos into generative latent representations and reconstructs them with perceptually or adversarially trained decoders~\citep{plvc2022,glcvideo2025,glvc2025,gvc1d2026}, while another incorporates diffusion models to synthesize plausible details~\citep{gnvcvd2026,s2vc2026,yoda2026,vortec2026}.
These perceptual gains can come at the cost of source fidelity, as illustrated by the content and structural deviations in Figure~\ref{fig:motivation}(a).
Balancing fidelity and realism under stringent bitrate constraints therefore remains challenging~\citep{blau2019rdp}.
As illustrated by the representative architectures in Figure~\ref{fig:interfaces}(b,c), existing designs also tie generation to codec-specific latent representations and conditional reconstruction pipelines, limiting reuse across codecs without redesign or retraining.
This raises the question: \emph{Can existing codecs achieve a better distortion--perception trade-off through simple, reusable adaptation?}

Video generation transports noise toward the natural video distribution, whereas compression maps source videos to a reconstruction distribution shaped by bitrate constraints.
Our insight is to connect these processes by adapting a pretrained video prior to transport native codec reconstructions back toward the natural video distribution.
For distortion-oriented codecs, these reconstructions retain basic source content and spatial layout, anchoring the transport to source videos so that refinement improves naturalness while maintaining source fidelity.
These reconstructions provide a shared pixel-space interface, decoupling generative refinement from codec-specific representations and reference updates, which supports lightweight adaptation and reuse across codecs.

Building on this insight, we introduce \textbf{Gen2-VC}, which adapts a pretrained video prior while preserving the codec's native reference update chain, as illustrated in Figure~\ref{fig:interfaces}(d).
With the codec, VAE, and generative backbone frozen, we train lightweight LoRA adapters~\citep{lora2022} in two stages using source video supervision: an intra-frame adapter (I-adapter) learns spatial refinement, then remains fixed while a predictive adapter (P-adapter) learns temporal refinement.
At inference, the I-adapter refines frames independently, and the P-adapter jointly refines overlapping windows using its previous outputs as context.
Each adapter call performs one near-terminal latent update.

Using Wan2.1-T2V-1.3B~\citep{wan2025}, Gen2-VC-DCVC-UF achieves a better distortion--perception trade-off at low bitrates. 
To our knowledge, it is the first generative video codec to achieve leading LPIPS~\citep{lpips2018} and DISTS~\citep{dists2022} performance while surpassing VTM-23.0~\citep{vtm} in PSNR and MS-SSIM~\citep{ms-ssim}, measured using BD-rates averaged over six datasets in this regime. 
Training-free transfer further delivers perceptual gains on DCVC-RT~\citep{dcvcrt2025}, ECM~\citep{ecm}, VTM~\citep{vtm}, and HM~\citep{hm}.
Given the same codec inputs, Gen2-VC also achieves lower average BD-rates than the evaluated video restoration methods.

Our contributions are summarized as follows:
\begin{itemize}
  \item We introduce a decoupled generative video compression framework that connects a shared video prior to learned and conventional codecs through native pixel-space reconstructions, preserving their bitstreams and reference update chains.

  \item We propose lightweight LoRA adaptation in spatial and temporal training stages, with one latent update per adapter call and the codec, VAE, and generative backbone frozen.

  \item Using one shared 1.3B video prior, we demonstrate improved rate--distortion--perception trade-offs at low bitrates across six datasets and training-free transfer of adapters trained only on DCVC-UF to other learned and conventional codecs.
\end{itemize}

\section{Related Work}
\label{sec:related_work}

\paragraph{Distortion-oriented video compression.}
Conventional codecs~\citep{hm,vtm,ecm} combine handcrafted prediction, transform coding, and entropy coding under rate--distortion optimization.
Among learned codecs, DVC~\citep{dvc2019} learns motion and residual coding, FVC~\citep{fvc2021} operates in feature space, and DCVC~\citep{dcvc2021} uses temporal context for conditional coding.
DCVC-RT~\citep{dcvcrt2025} and DCVC-UF~\citep{dcvcuf2026} further improve coding efficiency and throughput.
Gen2-VC uses their native reconstructions as a shared pixel-space interface for generative refinement, preserving existing coding pipelines.

\paragraph{Generative video compression.}
Generative codecs improve perceptual realism through two broad strategies. The first uses perceptual and adversarial training. PLVC~\citep{plvc2022} employs a recurrent conditional GAN, while GLC-Video~\citep{glcvideo2025} and GLVC~\citep{glvc2025} compress latents from a generative VQ-VAE and a pretrained video tokenizer, respectively. GVC1D~\citep{gvc1d2026} exploits compact one-dimensional tokens with short- and long-term context, while GVC-RT~\citep{gvcrt2026} learns tokenizer-aligned continuous latents for real-time coding. 
The second uses pretrained diffusion and flow models. GNVC-VD~\citep{gnvcvd2026} conditions VideoDiT on intermediate codec features to refine decoded latents. S$^2$VC~\citep{s2vc2026} conditions diffusion on semantic context.
YODA~\citep{yoda2026} applies temporal conditioning within its autoencoder and latent codec, while VoRTeC~\citep{vortec2026} combines latent coding with one-step flow refinement.
These designs tie generation to codec-specific representations and reconstruction pipelines. Gen2-VC instead uses native pixel-space reconstructions as a shared interface after independent codec decoding, as illustrated in Figure~\ref{fig:interfaces}(d).

\paragraph{Generative video restoration.}
Generative models support video detail recovery and super-resolution~\citep{he2024venhancer,upscaleavideo2024,videogigagan2025}.
SeedVR~\citep{seedvr} targets unknown real-world degradations, while SeedVR2~\citep{seedvr22026} enables efficient one-step restoration at high resolutions.
Generic restoration training may not match codec-induced distortions or the fidelity--perception balance required at low bitrates.
DiQP~\citep{diqp2025} targets quantization distortions introduced by conventional video compression.
Gen2-VC focuses on lightweight adaptation of a shared frozen video prior for low-bitrate rate--distortion--perception trade-offs, with training-free transfer across learned and conventional codecs.

\section{Methods}
\label{sec:methods}

\begin{figure}[!t]
  \centering
  \includegraphics[width=\linewidth]{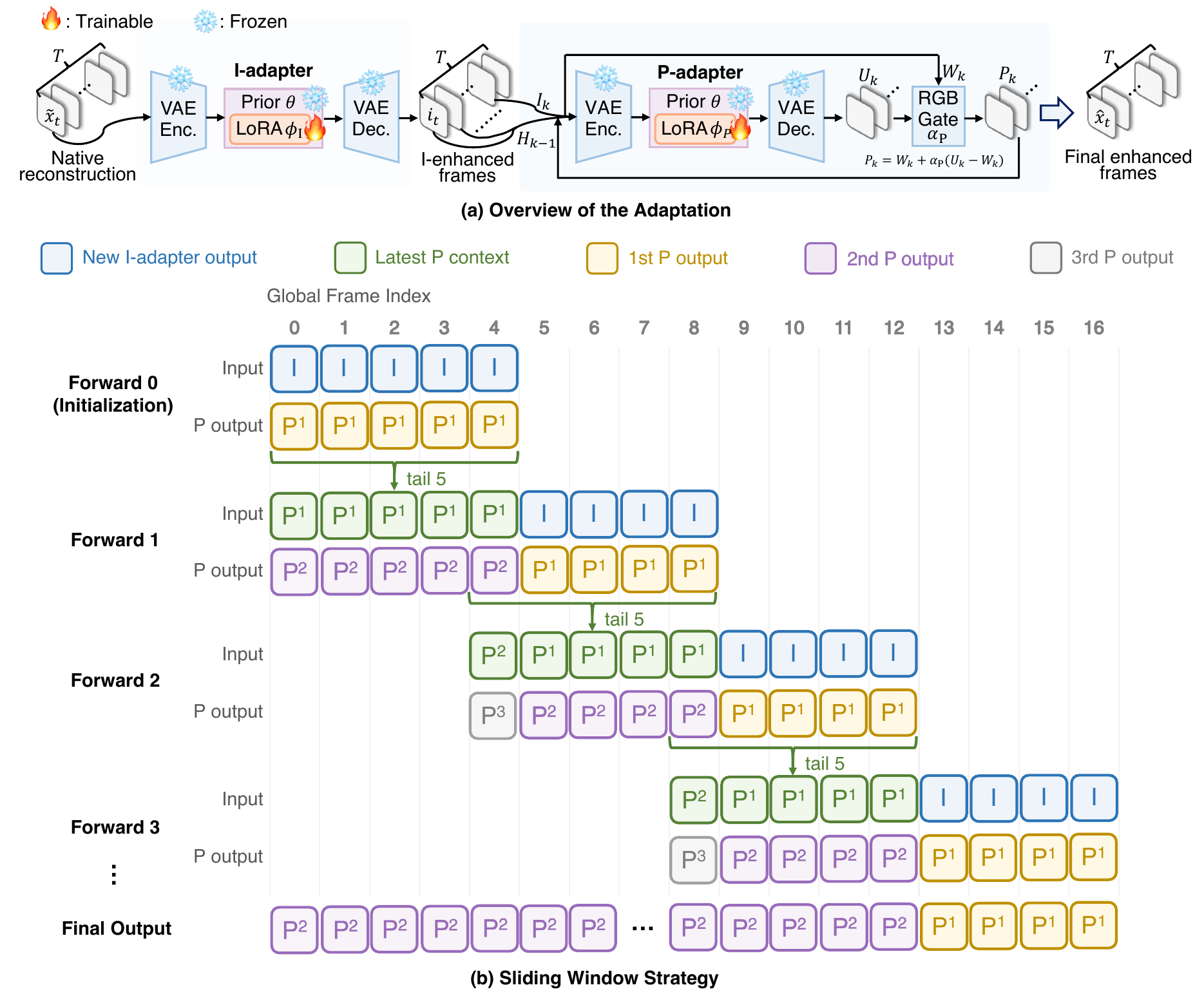}
  \caption{I-adapter and P-adapter in Gen2-VC. (a) The I-adapter restores individual frames before the P-adapter refines them jointly. (b) Sliding-window recurrence and output selection, shown with zero-based frame indices.}
  \label{fig:ip_adaptation}
\end{figure}

\paragraph{Overview.}
An existing codec encodes a source video $X=\{x_t\}_{t=1}^{T}$ into a bitstream, then decodes it to obtain the native reconstruction $\widetilde X=\{\tilde x_t\}_{t=1}^{T}$.
As illustrated in Figure~\ref{fig:ip_adaptation}(a), Gen2-VC first restores each frame of $\widetilde X$ independently with an intra-frame adapter (I-adapter), then jointly refines successive frames with a predictive adapter (P-adapter) to produce the final enhanced video $\hat X=\{\hat x_t\}_{t=1}^{T}$.
The P-adapter reuses its outputs as recurrent context within Gen2-VC, leaving the codec's bitstream and reference update process unchanged.
Both adapters share a frozen VAE and video prior and use separate LoRA parameters.

\subsection{Source-Anchored Spatial Adaptation}
\label{sec:spatial_adaptation}

Although entropy-constrained compression can remove texture and distort local structure, native reconstructions retain the source's core content and spatial layout, providing an anchor for detail recovery.
The I-adapter independently refines each native reconstruction $\tilde x_t$ into an I-enhanced frame $i_t$, regardless of its codec frame type.
Specifically, the frozen VAE encoder yields latent $z_t=\mathcal E(\tilde x_t)$.
The video prior performs one update at a fixed near-terminal timestep $\tau_\star$ without added noise:
\begin{equation}
  z_t^\ast=z_t-\beta_I\sigma_\star
    v_{\theta,\phi_I}(z_t;\tau_\star).
  \label{eq:spatial_update}
\end{equation}
The frozen VAE decoder then produces the I-enhanced frame $i_t=\mathcal D(z_t^\ast)$.
Here $\theta$ denotes the frozen backbone parameters, and $\phi_I$ denotes the LoRA~\citep{lora2022} parameters.
The predictor $v_{\theta,\phi_I}(\cdot;\tau_\star)$ provides the latent update signal under fixed conditioning shared across videos, requiring no source-specific auxiliary information.
The coefficient $\sigma_\star$ scales this prediction at $\tau_\star$, and $\beta_I$ is a fixed latent refinement strength for each operating point.

\begin{figure}[!t]
  \centering
  \includegraphics[width=\linewidth]{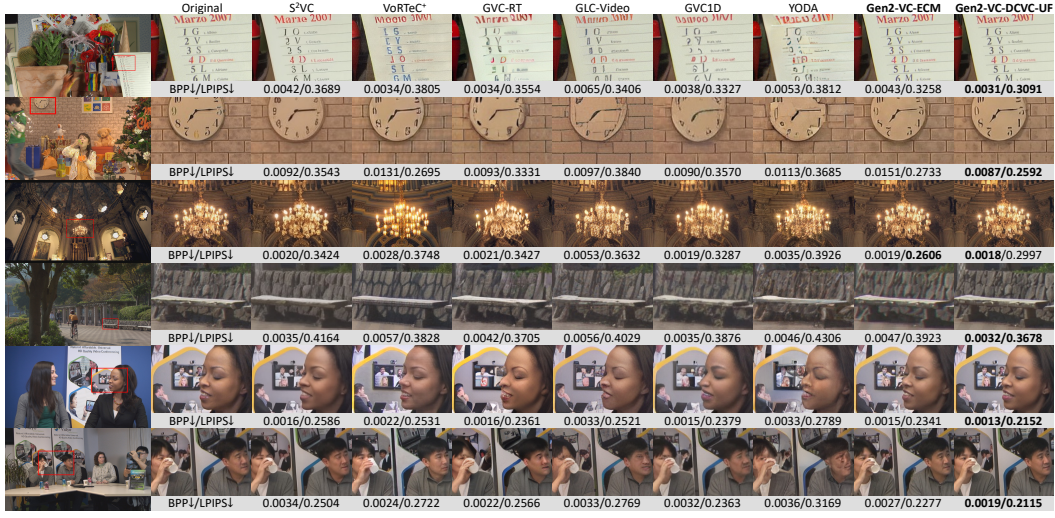}
  \caption{Gen2-VC better preserves source structure and details at low bitrates.
  Red boxes mark enlarged regions.
  BPP and LPIPS are reported below each reconstruction, with lower values preferred.}
  \label{fig:qualitative}
\end{figure}

\subsection{Recurrent Temporal Adaptation}
\label{sec:temporal_adaptation}

Independent frame refinement does not explicitly exploit temporal correspondence.
As illustrated in Figure~\ref{fig:ip_adaptation}(b), the P-adapter therefore jointly refines overlapping windows using its previous outputs as context.
Specifically, recurrent call $k\geq1$ receives $I_k=\{i_t\}_{t=t_k}^{t_k+L_{\mathrm{in}}-1}$, containing $L_{\mathrm{in}}$ new I-enhanced frames starting at source index $t_k$.
It combines these frames with history $H_{k-1}$ from the preceding gated window output $P_{k-1}$:
\begin{equation}
  H_{k-1}=\operatorname{tail}_{L_{\mathrm{tail}}}(P_{k-1}),\qquad
  W_k=\{H_{k-1},I_k\}.
  \label{eq:temporal_context}
\end{equation}
Here $\operatorname{tail}_{L_{\mathrm{tail}}}(\cdot)$ extracts the last $L_{\mathrm{tail}}$ frames, and braces denote temporal concatenation with history first, giving $L_{\mathrm{win}}=L_{\mathrm{tail}}+L_{\mathrm{in}}$ frames.
An initial call establishes history from I-enhanced frames.

The frozen VAE encoder yields latent $z_k^P=\mathcal E(W_k)$.
With LoRA parameters $\phi_P$ and fixed strength $\beta_P$, the video prior performs one update at $\tau_\star$.
The frozen VAE decoder produces a candidate window $U_k$, which the RGB gate $\alpha_{P,k}$ blends with $W_k$ to obtain the complete output $P_k$:
\begin{equation}
  \begin{aligned}
    U_k&=\mathcal D\!\left(z_k^P-\beta_P\sigma_\star
      v_{\theta,\phi_P}(z_k^P;\tau_\star)\right),\\
    P_k&=W_k+\alpha_{P,k}\odot(U_k-W_k).
  \end{aligned}
  \label{eq:temporal_update}
\end{equation}
Here $\odot$ denotes elementwise multiplication, and $\alpha_{P,k}$ controls refinement strength with first-visit and revisit scalars in $(0,1]$, shared across pixels and windows.
Both $U_k$ and $P_k$ retain all $L_{\mathrm{win}}$ frames of $W_k$.
The gated window's tail forms the next history $H_k$, while overlapping calls provide successive estimates for each frame.
Denoting frame $t$'s $j$-th P-adapter estimate by $p_t^{(j)}$, we select $\hat x_t=p_t^{(2)}$ when available and $p_t^{(1)}$ otherwise, as illustrated in Figure~\ref{fig:ip_adaptation}(b).
This finalizes $L_{\mathrm{out}}=L_{\mathrm{in}}$ new enhanced frames per call in steady state.
More details are provided in Appendix~\ref{app:training_details}.

\subsection{Staged Optimization}
\label{sec:staged_optimization}

We train the adapters in two stages.
First, we train the I-adapter to learn spatial refinement.
We then freeze the trained I-adapter and train the P-adapter to learn temporal refinement.
Throughout both stages, the codec, VAE, and pretrained backbone remain frozen.
For an enhanced training frame $y_t$, either $i_t$ or $p_t^{(j)}$, and the corresponding source frame $x_t$, the shared training objective is
\begin{equation}
  \mathcal L(y_t,x_t)
  =\lambda_1\lVert y_t-x_t\rVert_1
   +\lambda_{\mathrm{per}}\,\mathrm{LPIPS}(y_t,x_t)
   +\lambda_{\mathrm{adv}}\mathcal L_{\mathrm{adv}}.
  \label{eq:training_objective}
\end{equation}
The $\ell_1$ term constrains source fidelity, LPIPS~\citep{lpips2018} measures perceptual differences, and the adversarial loss encourages natural appearance.
For the I-adapter, we apply Eq.~(\ref{eq:training_objective}) independently to each frame with $\lambda_{\mathrm{adv}}=0.1$.
For the P-adapter, we average this loss over all frames in each window with $\lambda_{\mathrm{adv}}=0.0$ and detach history between calls.

\section{Experiments}
\label{sec:experiments}

\begin{figure}[!t]
  \centering
  \includegraphics[width=\linewidth]{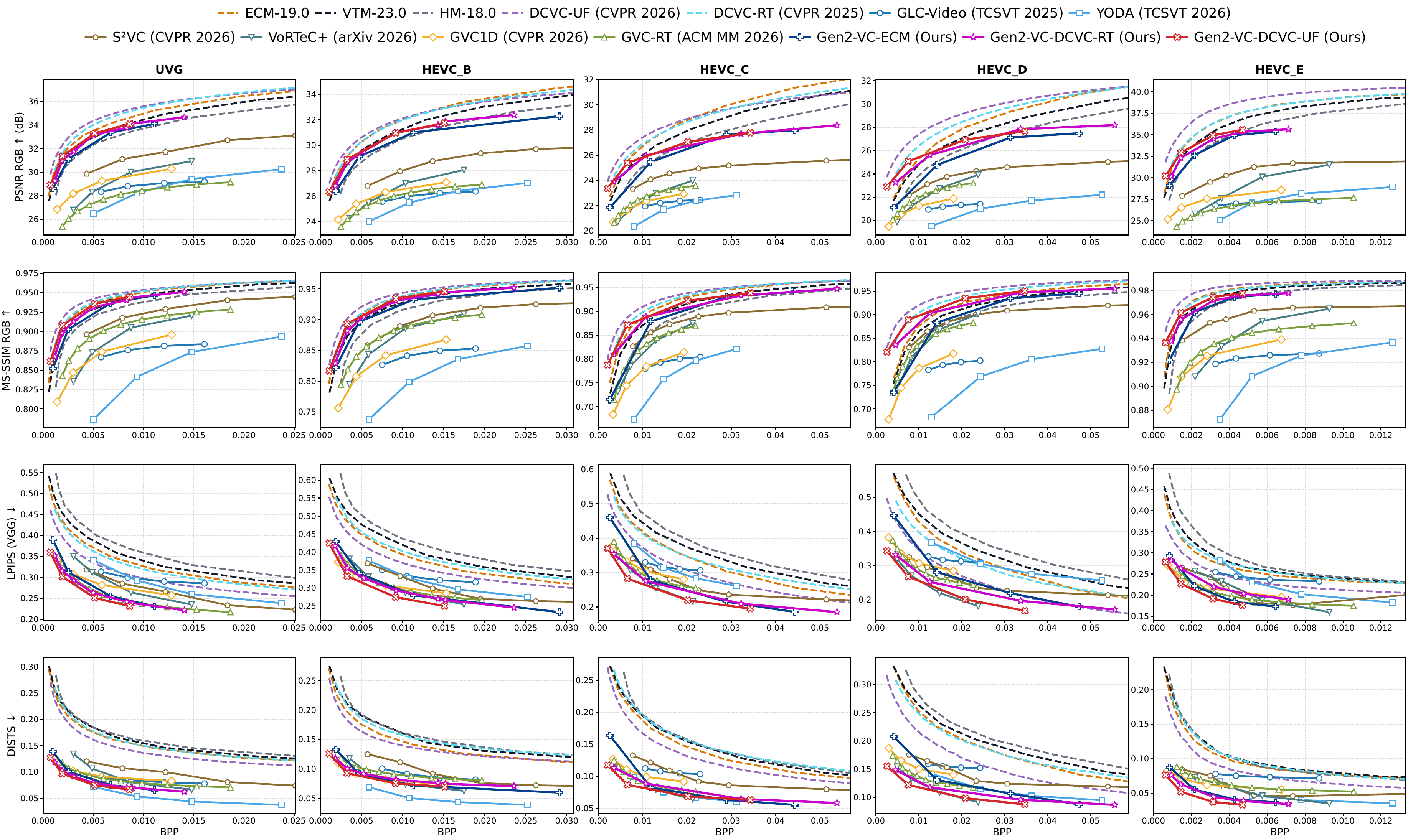}
  \caption{Rate--distortion--perception comparison on UVG and HEVC classes B/C/D/E.
  Higher is better for PSNR and MS-SSIM, while lower is better for LPIPS and DISTS.
  Bitrate is reported in bits per source pixel (BPP).}
  \label{fig:rd_primary}
\end{figure}

\subsection{Experimental Protocol}
\label{sec:experimental_protocol}
\paragraph{Training Details.}
We use Wan2.1-T2V-1.3B~\citep{wan2025} as the frozen video prior and apply LoRA only to its self-attention Q/K/V/O projections.
Both adapters use LoRA rank and scaling $(r,\alpha_{\mathrm{LoRA}})=(16,32)$ by default, each adding $0.42\%$ of the backbone parameters per operating point.
We first train the I-adapter on ImageNet~\citep{imagenet} and OpenImages~\citep{kuznetsova2020open} at variable resolutions of $512\times512$--$1024\times1024$, then the P-adapter on Vimeo~\citep{xue2019video} and OpenVid-HD~\citep{nan2025openvid} at $256\times256$--$720\times720$.
Training uses only DCVC-UF~\citep{dcvcuf2026} reconstructions. For training-free transfer, we select among the four I/P-adapter pairs by the codec's quantization parameter, keeping weights and refinement settings fixed.
Both stages use AdamW~\citep{AdamW2019} on two NVIDIA RTX PRO 6000 GPUs, with a batch size of $16$ and a learning rate decaying from $10^{-4}$ to $10^{-6}$.

\paragraph{Testing Details.}
Following the DCVC series~\citep{dcvc2021,dcvcrt2025,dcvcuf2026}, we evaluate the first $96$ frames of each sequence in UVG~\citep{uvg}, MCL-JCV~\citep{MCL-JCV}, and HEVC classes B/C/D/E~\citep{HEVC_Range_Extensions}, with intra-period $-1$.
We convert YUV inputs to RGB using BT.709, following JPEG AI~\citep{jpegai}, and compute PSNR, MS-SSIM~\citep{ms-ssim}, VGG-based LPIPS~\citep{lpips2018}, and DISTS~\citep{dists2022} in RGB space.
BD-rate~\citep{bdrate} uses PCHIP interpolation over the common quality range without extrapolation, where positive values indicate bitrate increases and negative values indicate bitrate savings at matched quality.
Appendix~\ref{app:additional_quantitative} reports patch-based FID~\citep{fid} and KID~\citep{kid} following GLC-Video~\citep{glcvideo2025}, together with MUSIQ~\citep{musiq2021}, NIQE~\citep{niqe2013}, and CLIP-IQA~\citep{clipiqa2023}.

\paragraph{Baselines.}
Distortion-oriented baselines comprise HM-18.0~\citep{hm}, VTM-23.0~\citep{vtm}, and ECM-19.0~\citep{ecm} under low-delay configurations, together with the learned codecs DCVC-RT~\citep{dcvcrt2025} and DCVC-UF-HTL~\citep{dcvcuf2026} (DCVC-UF hereafter).
Generative baselines include the VAE/tokenizer-based codecs GLC-Video~\citep{glcvideo2025}, GVC1D~\citep{gvc1d2026}, and GVC-RT~\citep{gvcrt2026}, and the diffusion-based codecs YODA~\citep{yoda2026}, S$^2$VC~\citep{s2vc2026}, and VoRTeC+~\citep{vortec2026}.
We evaluate restoration models SeedVR2~\citep{seedvr22026} with wavelet color correction, DiQP~\citep{diqp2025}, and VEnhancer-v2~\citep{he2024venhancer} on native DCVC-RT/DCVC-UF reconstructions.

\subsection{Experimental Results}
\begin{table}[!t]
\centering
\caption{BD-rate (\%) $\downarrow$ relative to VTM-23.0, averaged equally over UVG, MCL-JCV, HEVC-B, HEVC-C, HEVC-D, and HEVC-E. Missing coverage is shown as ``--''.}
\label{tab:bd_rate_summary}
\begingroup
\fontsize{9}{10}\selectfont
\setlength{\tabcolsep}{4pt}
\renewcommand{\arraystretch}{1.0}
\begin{tabular*}{\linewidth}{@{\extracolsep{\fill}}lrrrr@{}}
\toprule
Method & PSNR & MS-SSIM & LPIPS & DISTS \\
\midrule
\rowcolor{tablegray}
\multicolumn{5}{c}{\textit{Distortion-oriented video codec}} \\
HM-18.0 & $38.32$ & $43.93$ & $29.25$ & $11.55$ \\
VTM-23.0 (Anchor) & $0.00$ & $0.00$ & $0.00$ & $0.00$ \\
ECM-19.0 & $-20.95$ & $-19.05$ & $-19.77$ & $-15.28$ \\
DCVC-RT & \underline{$-21.41$} & $-22.69$ & $-18.42$ & $-1.11$ \\
DCVC-UF & $\mathbf{-37.16}$ & $\mathbf{-40.04}$ & $-49.41$ & $-36.39$ \\
\midrule
\rowcolor{tablegray}
\multicolumn{5}{c}{\textit{VAE/tokenizer-based video codec}} \\
GLC-Video & -- & $365.05$ & $-56.27$ & $-78.87$ \\
GVC1D & $403.37$ & $199.97$ & $-81.91$ & $-91.14$ \\
GVC-RT & $816.32$ & $123.22$ & \underline{$-81.99$} & $-89.38$ \\
\midrule
\rowcolor{tablegray}
\multicolumn{5}{c}{\textit{Diffusion-based video codec}} \\
YODA & $984.31$ & $703.65$ & $-57.14$ & $-90.64$ \\
S$^2$VC & $307.75$ & $126.36$ & $-65.30$ & $-77.58$ \\
VoRTeC+ & $409.33$ & $181.60$ & $-80.13$ & $-90.79$ \\
\midrule
\rowcolor{tablegray}
\multicolumn{5}{c}{\textit{Video restoration methods}} \\
SeedVR2 + DCVC-RT & -- & $360.94$ & $-16.13$ & $-33.16$ \\
VEnhancer-v2 + DCVC-RT & -- & -- & $4.91$ & $-28.04$ \\
DiQP + DCVC-RT & $55.37$ & $16.33$ & $37.95$ & $57.19$ \\
SeedVR2 + DCVC-UF & -- & $164.70$ & $-44.51$ & $-63.32$ \\
VEnhancer-v2 + DCVC-UF & -- & -- & $-30.09$ & $-48.18$ \\
DiQP + DCVC-UF & $5.60$ & $-18.74$ & $-9.16$ & $8.71$ \\
\midrule
\rowcolor{tablegray}
\multicolumn{5}{c}{\textit{Gen2-VC variants}} \\
Gen2-VC-HM & $94.00$ & $78.93$ & $-21.26$ & $-48.06$ \\
Gen2-VC-VTM & $50.77$ & $29.59$ & $-66.33$ & $-82.43$ \\
Gen2-VC-ECM & $28.63$ & $7.08$ & $-78.33$ & $-90.84$ \\
Gen2-VC-DCVC-RT & $8.72$ & $-7.28$ & $-79.90$ & \underline{$-91.76$} \\
Gen2-VC-DCVC-UF & $-12.11$ & \underline{$-28.03$} & $\mathbf{-86.65}$ & $\mathbf{-94.24}$ \\
\bottomrule
\end{tabular*}
\endgroup
\end{table}

\paragraph{Quantitative Evaluation.}
Figure~\ref{fig:rd_primary} shows rate--distortion--perception curves.
Gen2-VC-DCVC-UF substantially outperforms the evaluated generative codecs in PSNR/MS-SSIM across all datasets.
On six-dataset averages at low bitrates in Table~\ref{tab:bd_rate_summary}, it saves $86.65\%$/$94.24\%$ bitrate at matched LPIPS/DISTS and $12.11\%$/$28.03\%$ at matched PSNR/MS-SSIM relative to VTM-23.0.
To our knowledge, it is the first generative video codec to combine state-of-the-art LPIPS/DISTS performance with PSNR/MS-SSIM gains over VTM-23.0 on this six-dataset average.
Training-free transfer of the same DCVC-UF-trained adapters to DCVC-RT, ECM, VTM, and HM substantially lowers mean LPIPS/DISTS BD-rates without modifying their bitstreams or reference update chains, supporting reusable adaptation across learned and conventional codecs.

Using the same codec inputs, Gen2-VC also outperforms SeedVR2, DiQP, and VEnhancer-v2 in the available full-reference comparisons in Table~\ref{tab:bd_rate_summary}.
Appendices~\ref{app:additional_quantitative} and~\ref{app:per_dataset_bd_rate} provide additional metrics and per-dataset BD-rates.
SeedVR2 and VEnhancer-v2 attain favorable MUSIQ/CLIP-IQA scores despite weaker full-reference performance, suggesting a mismatch between restoration objectives and low-bitrate fidelity--perception requirements.

\paragraph{Qualitative Evaluation.}
Figures~\ref{fig:motivation}(a) and~\ref{fig:qualitative} illustrate why fidelity and perceptual quality should be assessed jointly at low bitrates.
DCVC-UF can oversmooth details despite strong PSNR/MS-SSIM, whereas GVC1D and YODA can distort source content despite perceptual gains.
Gen2-VC-DCVC-UF preserves structure and details more faithfully at the same or lower bitrates.
Our user study in Appendix~\ref{app:user_study} further supports Gen2-VC's gains in subjective quality.
Appendix~\ref{app:additional_visual} provides further codec and restoration comparisons.

\paragraph{Complexity Analysis.}
Table~\ref{tab:complexity} profiles I/P separately, each adding $5.90$M trainable parameters to the shared frozen VAE and backbone.
At every tested resolution, both stages have higher FPS than S$^2$VC, VoRTeC+, and all three restoration methods, fewer MACs than VEnhancer-v2, and less peak GPU memory than SeedVR2.
These component-level comparisons follow the profiling scopes detailed in Appendix~\ref{app:complexity_protocol}, which also reports the complete DCVC-UF-based decoding pipeline.
The full pipeline outpaces VoRTeC+ and the restoration methods on one GPU and supports multi-GPU pipelining by keeping refinement outside codec reference updates.
Gen2-VC thus combines parameter-efficient adaptation with competitive enhancement throughput while preserving native encoding efficiency, with all additional computation confined to decoder-side refinement.

\begin{table}[!t]
\centering
\caption{Computational complexity on one A100 80\,GB GPU. Video restoration methods and Gen2-VC report enhancement-only costs, with I- and P-adapters profiled separately. FPS: frames per second. TMAC/frame: Tera multiply-accumulate operations/frame. Memory: peak allocated/reserved (A/R) GPU memory, in GiB. Params.: total (trainable) parameters of default configurations, in M. ``--'' denotes the encoding FPS decided by the selected codec.}
\label{tab:complexity}
\begingroup
\fontsize{9}{10}\selectfont
\setlength{\tabcolsep}{3.5pt}
\renewcommand{\arraystretch}{0.90}
\begin{tabular*}{\linewidth}{@{\extracolsep{\fill}}llrrrrrr@{}}
\toprule
 & & \multicolumn{2}{c}{FPS $\uparrow$} & \multicolumn{2}{c}{TMAC/frame $\downarrow$} & Memory $\downarrow$ & Params. $\downarrow$ \\
\cmidrule(lr){3-4}\cmidrule(lr){5-6}
Method & Res. & Enc. & Dec. & Enc. & Dec. & A/R & Total (Train.) \\
\midrule
\rowcolor{tablegray}
\multicolumn{8}{c}{\textit{Diffusion-based video codec}} \\
YODA & 240p & 3.958 & 5.011 & 2.419 & 1.601 & 9.85/10.49 & 2055.13 \\
 & 480p & 2.860 & 3.920 & 4.469 & 2.951 & 11.64/12.93 &  \\
 & 720p & 1.536 & 2.292 & 10.279 & 6.776 & 16.69/21.37 &  \\
 & 1080p & 0.780 & 1.230 & 21.813 & 14.370 & 26.73/39.32 &  \\
\addlinespace[1pt]
S$^2$VC & 240p & 4.857 & 5.600 & 0.918 & 0.859 & 7.32/8.09 & 1842.29 \\
 & 480p & 2.148 & 2.483 & 3.532 & 3.310 & 8.25/9.41 &  \\
 & 720p & 0.987 & 1.113 & 8.650 & 8.138 & 9.94/12.50 &  \\
 & 1080p & 0.404 & 0.439 & 20.490 & 19.403 & 13.31/18.61 &  \\
\addlinespace[1pt]
VoRTeC+ & 240p & 9.341 & 6.312 & 0.344 & 0.893 & 9.24/10.01 & 1802.72 \\
 & 480p & 0.977 & 0.482 & 1.288 & 3.551 & 14.62/17.54 &  \\
 & 720p & 0.476 & 0.240 & 3.011 & 9.271 & 24.18/29.76 &  \\
 & 1080p & 0.188 & 0.097 & 6.561 & 24.290 & 43.20/56.00 &  \\
\midrule
\rowcolor{tablegray}
\multicolumn{8}{c}{\textit{Video restoration methods}} \\
SeedVR2 & 240p & -- & 0.368 & -- & 2.014 & 36.36/36.74 & 3642.12 \\
 & 480p & -- & 0.344 & -- & 8.028 & 39.94/40.55 &  \\
 & 720p & -- & 0.258 & -- & 18.586 & 46.87/49.11 &  \\
 & 1080p & -- & 0.194 & -- & 42.515 & 68.37/78.09 &  \\
\addlinespace[1pt]
DiQP & 240p & -- & 3.050 & -- & 0.589 & 1.48/2.15 & 79.37 \\
 & 480p & -- & 1.695 & -- & 1.178 & 1.48/2.15 &  \\
 & 720p & -- & 0.527 & -- & 3.534 & 1.48/2.15 &  \\
 & 1080p & -- & 0.261 & -- & 7.067 & 1.48/2.15 &  \\
\addlinespace[1pt]
VEnhancer-v2 & 240p & -- & 0.072 & -- & 145.383 & 19.55/78.62 & 2496.59 \\
 & 480p & -- & 0.072 & -- & 145.383 & 19.58/78.62 &  \\
 & 720p & -- & 0.071 & -- & 145.383 & 19.63/78.62 &  \\
 & 1080p & -- & 0.093 & -- & 216.563 & 23.81/78.62 &  \\
\midrule
\rowcolor{tablegray}
\multicolumn{8}{c}{\textit{Gen2-VC variants}} \\
I-adapter & 240p & -- & 16.659 & -- & 2.071 & 3.27/5.42 & 1551.79 (5.90) \\
 & 480p & -- & 5.234 & -- & 8.205 & 3.29/12.56 &  \\
 & 720p & -- & 2.274 & -- & 20.268 & 3.34/24.20 &  \\
 & 1080p & -- & 0.887 & -- & 53.519 & 10.83/27.21 &  \\
\addlinespace[1pt]
P-adapter & 240p & -- & 12.807 & -- & 2.430 & 5.21/7.37 & 1551.79 (5.90) \\
 & 480p & -- & 3.553 & -- & 10.396 & 11.22/19.35 &  \\
 & 720p & -- & 1.477 & -- & 27.122 & 21.67/39.08 &  \\
 & 1080p & -- & 0.557 & -- & 77.546 & 45.25/69.17 &  \\
\bottomrule
\end{tabular*}
\par\vspace{3pt}
\endgroup
\end{table}

\subsection{Ablation Study}
\label{sec:ablation}
\begin{table}[!t]
\centering
\caption{Stage and rank ablations on HEVC-E using DCVC-UF and four adapter operating points. BD-rate (\%) $\downarrow$ relative to VTM-23.0.}
\label{tab:ablation_study}
\begingroup
\fontsize{9}{10}\selectfont
\setlength{\tabcolsep}{3pt}
\renewcommand{\arraystretch}{1.02}
\begin{tabular*}{\linewidth}{@{\extracolsep{\fill}}llrrrrr@{}}
\toprule
Variant & P output / rank & Total LoRA (M) & PSNR & MS-SSIM & LPIPS & DISTS \\
\midrule
\rowcolor{tablegray}
\multicolumn{7}{c}{\textit{(a) Stages and output selection}} \\
Native DCVC-UF & -- & -- & \textbf{-52.88} & \textbf{-46.08} & -53.05 & -42.83 \\
\textit{I-only} & -- & 5.90 & -7.99 & -18.06 & -88.35 & -96.52 \\
\textit{I+P} & \textit{First} & 11.80 & -8.95 & -18.70 & \underline{-88.73} & \underline{-96.75} \\
\textit{I+P} & \textit{Second} & 11.80 & -12.89 & -22.51 & \textbf{-92.06} & \textbf{-97.47} \\
\textit{pixel-P} (no I) & \textit{First} & 11.80 & \underline{-43.01} & \underline{-38.00} & -78.36 & -87.65 \\
\textit{pixel-P} (no I) & \textit{Second} & 11.80 & -37.09 & -34.09 & -81.04 & -89.55 \\
\midrule
\rowcolor{tablegray}
\multicolumn{7}{c}{\textit{(b) I-adapter rank}} \\
\textit{I-only} & $r_I=8$ & 2.95 & \underline{-10.61} & \textbf{-22.51} & -84.48 & \textbf{-96.62} \\
\textit{I-only} (default) & $r_I=16$ & 5.90 & -7.99 & -18.06 & \underline{-88.35} & -96.52 \\
\textit{I-only} & $r_I=32$ & 11.80 & \textbf{-10.89} & \underline{-21.01} & \textbf{-92.98} & \underline{-96.59} \\
\midrule
\rowcolor{tablegray}
\multicolumn{7}{c}{\textit{(c) P-adapter rank: fixed $r_I=16$, Second outputs}} \\
\textit{I+P} (default) & $r_P=16$ & 11.80 & \textbf{-12.89} & \textbf{-22.51} & -92.06 & -97.47 \\
\textit{I+P} & $r_P=32$ & 17.69 & \underline{-12.72} & \underline{-21.47} & \textbf{-92.20} & \underline{-97.48} \\
\textit{I+P} & $r_P=64$ & 29.49 & \underline{-12.72} & -21.46 & \underline{-92.18} & \textbf{-97.51} \\
\bottomrule
\end{tabular*}
\endgroup
\end{table}

We select settings by jointly considering LPIPS/DISTS and LoRA parameter cost,
while retaining PSNR/MS-SSIM gains over VTM-23.0 for BD-rate averaged over six datasets.

\paragraph{Effects of the P-Stage and Output Selection.}
We test recurrent refinement by comparing \textit{I-only}, which applies the I-adapter framewise, with \textit{I+P}, which adds the P-adapter.
Across overlapping windows, \textit{First} retains each frame's first P estimate, while \textit{Second} uses its second when available and otherwise the first, following Figure~\ref{fig:ip_adaptation}(b).
In Table~\ref{tab:ablation_study}(a), \textit{Second} improves all four BD-rates over \textit{I-only} and \textit{First}, lowering LPIPS/DISTS from \textit{I-only}'s $-88.35\%$/$-96.52\%$ to $-92.06\%$/$-97.47\%$.
FloLPIPS~\citep{FloLPIPS2022} results in Appendix~\ref{app:additional_ablation} and cross-frame examples in Appendix~\ref{app:visual_stages} further support these temporal refinement gains over the evaluated 96-frame clips.

\paragraph{Effectiveness of Staged Adaptation.}
To test adapter separation, \textit{pixel-P} uses one rank-32 adapter trained first on single frames and then on multiple frames to refine native reconstructions directly.
Its single- and multi-frame stages share the hardware and dataset mixtures used to train the I- and P-adapters, respectively, and both pipelines use $11.80$M LoRA parameters.
With \textit{Second} outputs, \textit{I+P} improves LPIPS/DISTS BD-rates from \textit{pixel-P}'s $-81.04\%$/$-89.55\%$ to $-92.06\%$/$-97.47\%$ in Table~\ref{tab:ablation_study}(a).
Under fixed GPU memory, multi-frame training uses lower resolutions than single-frame training, as described in Section~\ref{sec:experimental_protocol}.
Appendix~\ref{app:additional_ablation} shows that \textit{pixel-P} can lose perceptual quality at higher test resolutions despite spatial pretraining, consistent with this train--test mismatch.
\textit{I+P} preserves higher-resolution spatial refinement through the fixed I-adapter's outputs, allowing the P-adapter to focus on temporal modeling and supporting parameter-efficient staged adaptation.

\paragraph{LoRA Rank and Adaptation Capacity.}
Table~\ref{tab:ablation_study}(b) compares I-adapter ranks 8, 16, and 32 under a common training recipe.
Considering LPIPS and DISTS jointly, we select $r_I=16$ over rank 8.
Rank 32 offers further gains but doubles the LoRA parameters.
Under the same GPU memory budget for training, $r_P=16$ offers the best PSNR/MS-SSIM and comparable LPIPS/DISTS in Table~\ref{tab:ablation_study}(c), with $11.80$M total LoRA parameters versus $17.69$M/$29.49$M for ranks 32/64.
Six-dataset averages in Appendix~\ref{app:rank_ablation} further support $r_P=16$ as our default.

\section{Conclusion}
\label{sec:conclusion}

We presented Gen2-VC, a decoupled framework that connects a pretrained video prior to existing codecs through native pixel-space reconstructions, anchoring refinement to source content while preserving bitstreams and native reference update chains.
Lightweight LoRA adapters learn spatial and temporal refinement in stages, with one latent update per adapter call and the codec, VAE, and backbone frozen.
Experiments with one shared Wan2.1-T2V-1.3B prior demonstrate improved rate--distortion--perception trade-offs at low bitrates across six datasets.
Adapters trained only on DCVC-UF further deliver perceptual gains on other learned and conventional codecs through training-free transfer.
Gen2-VC thus provides a practical route to incorporate generative refinement into existing video compression systems without redesigning their coding pipelines.

\clearpage
\subsection*{AI use statement}
The authors confirm that no AI-assisted tools were used in the preparation of this manuscript.

\subsection*{Reproducibility statement}

Sections~\ref{sec:methods} and~\ref{sec:experiments} describe the method and
experimental protocol.
Appendices~\ref{app:training_details}, \ref{app:conventional_codec_testing},
and~\ref{app:complexity_protocol} provide implementation details, evaluation
settings, and profiling procedures.
We will publicly release the code and model weights upon acceptance.

\clearpage
\appendix
\appendixfloatplacement
\section{Appendix}
\label{sec:appendix}

The appendix is organized as follows.
\begin{description}
  \setlength{\itemsep}{1pt}
  \setlength{\parskip}{0pt}
  \item[Appendix~\ref{app:training_details}] Implementation, training, and recurrent inference.
  \item[Appendix~\ref{app:conventional_codec_testing}] Codec configurations, color conversion, and metric implementations.
  \item[Appendix~\ref{app:additional_quantitative}] Additional quality metrics.
  \item[Appendix~\ref{app:extended_comparisons}] Extended codec and restoration comparisons.
  \item[Appendix~\ref{app:additional_ablation}] Stage, output-selection, and LoRA-rank ablations.
  \item[Appendix~\ref{app:user_study}] User-study interface, protocol, and preference results.
  \item[Appendix~\ref{app:complexity_protocol}] Profiling protocol, component costs, and full decoding pipelines.
  \item[Appendix~\ref{app:per_dataset_bd_rate}] Per-dataset BD-rates and aggregation rules.
  \item[Appendix~\ref{app:additional_visual}] Codec, stage, and restoration visual comparisons.
\end{description}

\subsection{Implementation and Training Details}
\label{app:training_details}

Gen2-VC takes native RGB reconstructions as its pixel-space interface and
uses the frozen VAE to encode and decode refinement latents.

\paragraph{Training Configuration.}
Table~\ref{tab:appendix_training_config} summarizes training settings, adapter sizes, and refinement controls.
The codec, VAE, and backbone remain frozen, as does the I-adapter during P-stage training.
LoRA targets self-attention Q/K/V/O projections in all 30 backbone blocks,
with no dropout. Rank comparisons appear in Appendix~\ref{app:rank_ablation}.

\par\medskip
\noindent\begin{minipage}{\linewidth}
  \centering
  \captionof{table}{Training settings, default adapter sizes, and refinement controls.}
  \label{tab:appendix_training_config}
  \small
  \setlength{\tabcolsep}{5pt}
  \renewcommand{\arraystretch}{1.12}
  \begin{tabular}{@{}p{0.33\linewidth}p{0.30\linewidth}p{0.30\linewidth}@{}}
    \toprule
    Setting & I-adapter & P-adapter \\
    \midrule
    Training data & ImageNet, OpenImages & Vimeo, OpenVid-HD \\
    Spatial resolution & $512\times512$--$1024\times1024$ & $256\times256$--$720\times720$ \\
    Training codec & \multicolumn{2}{l}{DCVC-UF} \\
    LoRA variables & $\phi_I$ & $\phi_P$ \\
    LoRA $(r,\alpha_{\mathrm{LoRA}})$ & $(16,32)$ & $(16,32)$ \\
    LoRA parameters & 5,898,240 ($0.42\%$) & 5,898,240 ($0.42\%$) \\
    $(\lambda_1,\lambda_{\mathrm{per}},\lambda_{\mathrm{adv}})$ & $(1.0,0.8,0.1)$ & $(1.0,0.8,0.0)$ \\
    Discriminator & PatchGAN (training only) & None \\
    Optimizer & \multicolumn{2}{l}{AdamW} \\
    Learning rate & \multicolumn{2}{l}{$10^{-4}$ decaying to $10^{-6}$} \\
    Batch size & \multicolumn{2}{l}{16} \\
    Training iterations & 6,000 & 1,000 \\
    Hardware & \multicolumn{2}{l}{Two NVIDIA RTX PRO 6000 GPUs} \\
    \midrule
    Refinement strength & $(1.00,0.85,0.70,0.60)$ & $(1.00,0.70,0.35,0.25)$ \\
    RGB gate (first/revisit) & -- & $0.02\,/\,0.09$ \\
    \midrule
    $(L_{\mathrm{win}},L_{\mathrm{tail}},L_{\mathrm{in}},L_{\mathrm{out}})$ & -- & $(9,5,4,4)$ \\
    Initial context & -- & Five I-enhanced frames \\
    Training inputs & Native reconstructions & Fixed I-enhanced frames \\
    Loss aggregation & Per frame & Mean over all window frames \\
    History gradients & -- & Detached between calls \\
    Inference & Framewise & Section~\ref{sec:temporal_adaptation}, Figure~\ref{fig:ip_adaptation}(b) \\
    \bottomrule
  \end{tabular}
\end{minipage}
\par\medskip

Each of the four DCVC-UF operating points has a separately trained I/P-adapter pair.
LoRA counts are per operating point, with percentages relative to the
1,418,996,800 frozen backbone parameters, excluding the VAE and codec.
Refinement strengths and RGB gates are set empirically before training to
control refinement at different bitrates and remain fixed during training and inference.
Strength tuples follow increasing bitrate, while the two RGB gates are shared across all four points.
Both controls transfer unchanged, without further tuning for individual
codecs, datasets, videos, or frames.
At inference, we select among the four adapter pairs using the target
codec's quantization parameter, keeping the selected weights
and all other inference settings fixed.

\paragraph{One-update Enhancement.}
To align conditioning with the video prior, we follow Wan2.1's official
inference example for the CFG scale of $6$ and the negative prompt.
Both adapters share an empty positive prompt and fixed, precomputed UMT5
embeddings across videos and bitrate points, without loading or running the
text encoder at inference.
CFG combines positive/negative predictions as $v=v_-+6(v_+-v_-)$.
Index $49$ (zero-based) of a 50-point Diffusers 0.39.0
\texttt{UniPCMultistepScheduler} schedule with 1,000 training timesteps and
flow shift $3$ gives $\sigma_\star\approx0.06040539$ and
$\tau_\star=\lfloor1000\sigma_\star\rfloor=60$.
Each call scales RGB inputs to $[-1,1]$, applies VAE posterior-mode encoding
and Wan's latent normalization, then performs one latent update without added
noise as in Eqs.~(\ref{eq:spatial_update}) and~(\ref{eq:temporal_update}), with the next noise level set to zero.
We invert the normalization, decode, then rescale and clip outputs to $[0,1]$.

\begin{quote}
\begin{minipage}{\linewidth}
\small
\noindent\textbf{Negative prompt:}
Bright tones, overexposed, static, blurred details, subtitles, style, works,
paintings, images, static, overall gray, worst quality, low quality, JPEG
compression residue, ugly, incomplete, extra fingers, poorly drawn hands,
poorly drawn faces, deformed, disfigured, misshapen limbs, fused fingers,
still picture, messy background, three legs, many people in the background,
walking backwards
\end{minipage}
\end{quote}

\FloatBarrier
\subsection{Testing Protocol and Metric Implementations}
\label{app:conventional_codec_testing}

\paragraph{Conventional codec configurations.}
Table~\ref{tab:conventional_codec_configs} lists the low-delay configurations.
Paths are relative to each codec's installation directory.

\par\medskip
\noindent\begin{minipage}{\linewidth}
\centering
\captionof{table}{Conventional codec executables and low-delay configurations.}
\label{tab:conventional_codec_configs}
\begingroup
\small
\setlength{\tabcolsep}{4pt}
\renewcommand{\arraystretch}{1.12}
\begin{tabular}{@{}>{\raggedright\arraybackslash}p{0.17\linewidth}>{\raggedright\arraybackslash}p{0.23\linewidth}>{\raggedright\arraybackslash}p{\dimexpr0.60\linewidth-4\tabcolsep\relax}@{}}
\toprule
Codec & Executable & Configuration file \\
\midrule
HM-18.0 & \texttt{TAppEncoder} & \path{cfg/encoder_lowdelay_main10.cfg} \\
VTM-23.0 & \texttt{EncoderApp} & \path{cfg/encoder_lowdelay_vtm.cfg} \\
ECM-19.0 & \texttt{EncoderApp} & \path{cfg/encoder_lowdelay_ecm.cfg} \\
\bottomrule
\end{tabular}
\endgroup
\end{minipage}
\par\medskip

\paragraph{Test command and evaluation.}
We encode the first $96$ frames of each planar 8-bit YUV~4:2:0 sequence at
its source resolution, setting \texttt{frame\_number=96} and
\texttt{intra\_period=-1} below. Select the configuration from
Table~\ref{tab:conventional_codec_configs} and use \texttt{TAppEncoder}
instead of \texttt{EncoderApp} for HM.
The launcher's default \texttt{frame\_rate} is $24$.

\begingroup
\small
\begin{verbatim}
EncoderApp -c {encoder_configuration.cfg} \
  -f {frame_number} -q {qp} --IntraPeriod={intra_period} \
  --InputFile={src_yuv} \
  --SourceWidth={width} --SourceHeight={height} \
  --FrameRate={frame_rate} --Level=6.2 --InputBitDepth=8 \
  --DecodingRefreshType=2 -b {output.bin} -o {enc.yuv}
\end{verbatim}
\endgroup

The \texttt{-b} option writes \path{output.bin}, and \texttt{-o} writes the
native YUV reconstruction \path{enc.yuv}.
Bitrate is $\mathrm{BPP}=8\lvert\texttt{output.bin}\rvert_{\mathrm{bytes}}/(96HW)$
for frame size $H\times W$.
Following Table~\ref{tab:rgb_conversion}, we convert \path{src_yuv} and
\path{enc.yuv} to RGB8 for evaluation.
The converted native frames also serve as Gen2-VC inputs.
Framewise metrics in Table~\ref{tab:metric_implementations} are averaged over the
$96$ frames within each sequence, then equally across sequences in each dataset.

\par\medskip
\noindent\begin{minipage}{\linewidth}
\centering
\captionof{table}{Source and native-reconstruction conversion to 8-bit RGB.}
\label{tab:rgb_conversion}
\begingroup
\small
\setlength{\tabcolsep}{4pt}
\renewcommand{\arraystretch}{1.12}
\begin{tabular}{@{}>{\raggedright\arraybackslash}p{0.20\linewidth}>{\raggedright\arraybackslash}p{\dimexpr0.80\linewidth-2\tabcolsep\relax}@{}}
\toprule
Step & Input convention and implementation \\
\midrule
Read and normalize & NumPy reads planar YUV~4:2:0. Source samples are 8-bit
and divided by $255$. Native reconstructions are read as 10-bit samples in
little-endian 16-bit words and divided by $1023$. \\
\addlinespace[3pt]
Upsample chroma & SciPy \texttt{ndimage.zoom}, with \texttt{order=0}, performs
nearest-neighbor $2\times$ upsampling to luma resolution. \\
\addlinespace[3pt]
Convert color & A custom NumPy implementation applies
full-range BT.709 with $(K_R,K_G,K_B)=(0.2126,0.7152,0.0722)$,
following the DCVC-RT transform for both source and reconstruction. \\
\addlinespace[3pt]
Quantize and save & \texttt{np.rint(np.clip(rgb,0,1)*255)} followed by
\texttt{uint8} conversion produces RGB8. Pillow reads and writes the RGB PNGs
used for evaluation. \\
\bottomrule
\end{tabular}
\endgroup
\end{minipage}
\par\medskip

\paragraph{Baseline availability and rates.}
GLVC~\citep{glvc2025} and GNVC-VD~\citep{gnvcvd2026} are excluded from
quantitative comparisons because their code was unavailable at evaluation time.
HM, VTM, ECM, DCVC-RT, DCVC-UF, YODA, and GVC-RT use actual bitstream rates,
whereas GLC-Video and GVC1D use entropy-model estimates.
S$^2$VC combines estimated I-frame latent rates, excluding hyperlatents,
with measured P-frame rANS bytes.
VoRTeC+ estimates latent/AIN rates, including tail padding but excluding
auxiliary metadata.
Gen2-VC and restoration methods retain their input codec's bitrate.

\paragraph{DiQP evaluation.}
We evaluate DiQP using its highest quantization setting, typically associated
with the lowest-bitrate end of its original codec range.

\paragraph{SeedVR2 color correction.}
\label{app:seedvr2_color_fix}
We apply the wavelet color correction supported by the official SeedVR2
implementation to its quantized outputs, using the corresponding native
reconstructions as color references.
At identical bitrates, Table~\ref{tab:seedvr2_color_fix} shows improved
six-dataset mean BD-rates for MS-SSIM, LPIPS, and DISTS on both codecs.
We therefore include color correction in both quality evaluation and timing. 
PSNR is omitted because its quality ranges do not overlap VTM-23.0.
BD-rate follows the common quality-sorting convention before PCHIP
integration, including for non-monotonic curves.

\par\medskip
\noindent\begin{minipage}{\linewidth}
\centering
\captionof{table}{SeedVR2 with and without wavelet color correction.
BD-rate (\%) $\downarrow$ relative to VTM-23.0, averaged equally over UVG,
MCL-JCV, HEVC-B, HEVC-C, HEVC-D, and HEVC-E.}
\label{tab:seedvr2_color_fix}
\small
\setlength{\tabcolsep}{5pt}
\renewcommand{\arraystretch}{1.08}
\begin{tabular*}{\linewidth}{@{\extracolsep{\fill}}llrrr@{}}
\toprule
Method & Color fix & MS-SSIM & LPIPS & DISTS \\
\midrule
SeedVR2 + DCVC-RT & Without & \underline{461.07} & \underline{-12.50} & \underline{-8.98} \\
SeedVR2 + DCVC-RT & With & \textbf{360.94} & \textbf{-16.13} & \textbf{-33.16} \\
\midrule
SeedVR2 + DCVC-UF & Without & \underline{234.79} & \underline{-41.77} & \underline{-61.46} \\
SeedVR2 + DCVC-UF & With & \textbf{164.70} & \textbf{-44.51} & \textbf{-63.32} \\
\bottomrule
\end{tabular*}
\end{minipage}
\par\medskip

\subsubsection{Metric Implementations}
\label{app:metric_implementations}
The main quantitative comparisons use saved, quantized RGB frames.
Table~\ref{tab:metric_implementations} records the libraries and explicit call
settings in the evaluation code. FID/KID pool patches from all frames and
sequences at each dataset/rate point, rather than averaging framewise scores.

\noindent FID/KID use $64\times64$ patches when the dataset's minimum frame
dimension is below $256$, and $256\times256$ otherwise. Patches are extracted
with \texttt{torch.nn.functional.unfold} at stride equal to patch size, using
GLC's two-offset rule (\texttt{split\_patch\_num=2}) when the shifted grid fits.
NIQE retries numerical linear-algebra failures with the same PyIQA
implementation on CPU. Its averages use finite frames and then equally
weight sequences with a defined mean.

\par\medskip
\noindent\begin{minipage}{\linewidth}
\centering
\captionof{table}{Metric libraries and evaluation settings.}
\label{tab:metric_implementations}
\begingroup
\small
\setlength{\tabcolsep}{4pt}
\renewcommand{\arraystretch}{1.12}
\begin{tabular}{@{}>{\raggedright\arraybackslash}p{0.13\linewidth}>{\raggedright\arraybackslash}p{0.35\linewidth}>{\raggedright\arraybackslash}p{\dimexpr0.52\linewidth-4\tabcolsep\relax}@{}}
\toprule
Metric & Implementation & Key settings \\
\midrule
PSNR & NumPy (custom implementation) &
Float64 RGB MSE on $[0,1]$, $-10\log_{10}\mathrm{MSE}$. Perfect frames use a
$120$\,dB cap. \\
\addlinespace[3pt]
MS-SSIM & \texttt{pytorch-msssim} &
RGB in $[0,1]$, unit data range, per-frame scores. \\
\addlinespace[3pt]
LPIPS & \texttt{lpips} &
VGG backbone, RGB rescaled to $[-1,1]$. \\
\addlinespace[3pt]
FloLPIPS & Official FloLPIPS implementation &
AlexNet backbone and FP32 PWC optical flow. Average 95 adjacent-frame pairs per
96-frame sequence, then equally average sequences within each dataset. \\
\addlinespace[3pt]
DISTS & \texttt{DISTS-pytorch} &
RGB in $[0,1]$, evaluation mode. \\
\addlinespace[3pt]
FID / KID & TorchMetrics with \texttt{torch-fidelity} &
Default constructors, uint8 GLC-style patches. KID reports the mean and uses
PyTorch seed $0$ per dataset/rate point. \\
\addlinespace[3pt]
MUSIQ & PyIQA, \texttt{musiq} &
Reconstructed RGB in $[0,1]$, default KonIQ checkpoint. \\
\addlinespace[3pt]
NIQE & PyIQA, \texttt{niqe} &
Reconstructed RGB in $[0,1]$, default model. Undefined values are excluded
with coverage recorded. \\
\addlinespace[3pt]
CLIP-IQA & PyIQA, \texttt{clipiqa} &
Reconstructed RGB in $[0,1]$, default CLIP RN50 weights. \\
\addlinespace[3pt]
BD-rate & \texttt{bjontegaard} $1.3.0$ &
PCHIP over the common quality interval, without extrapolation.
MS-SSIM uses raw values, without dB conversion. \\
\bottomrule
\end{tabular}
\endgroup
\end{minipage}
\par\medskip

\FloatBarrier
\subsubsection{Additional Quality Metrics}
\label{app:additional_quantitative}
\begin{figure}[!htbp]
  \centering
  \includegraphics[width=\linewidth]{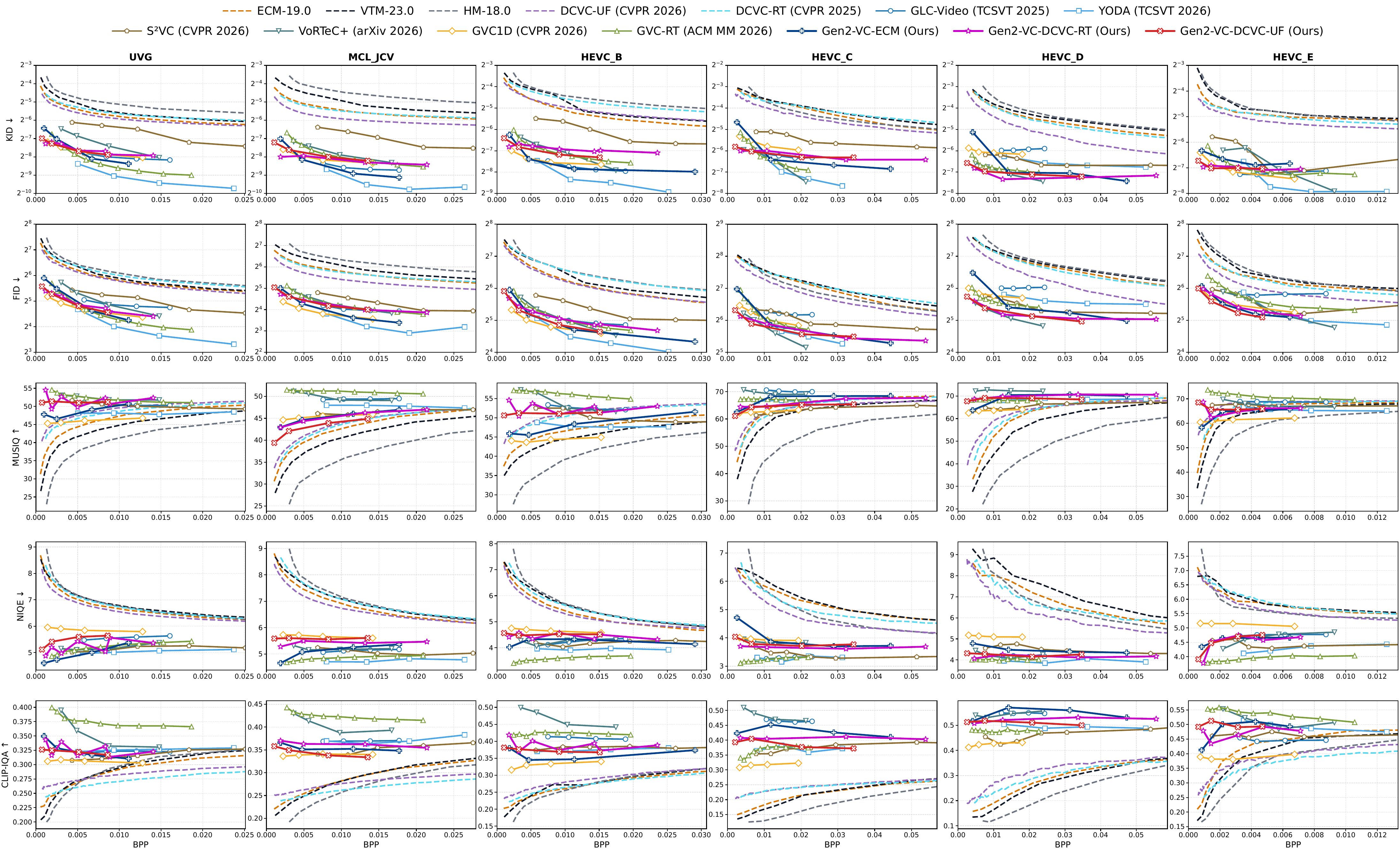}
  \caption{FID, KID, MUSIQ, NIQE, and CLIP-IQA comparisons with distortion-oriented and generative codecs.}
  \label{fig:rd_additional_metrics}
\end{figure}

Figure~\ref{fig:rd_additional_metrics} shows that Gen2-VC substantially
reduces FID and KID relative to native codec reconstructions across all six
datasets. Lower NIQE and generally higher MUSIQ/CLIP-IQA at low bitrates
further support its perceptual gains beyond LPIPS and DISTS.

\FloatBarrier

\subsubsection{Extended Codec and Restoration Comparisons}
\label{app:extended_comparisons}
\begin{figure}[!htbp]
  \centering
  \includegraphics[width=\linewidth]{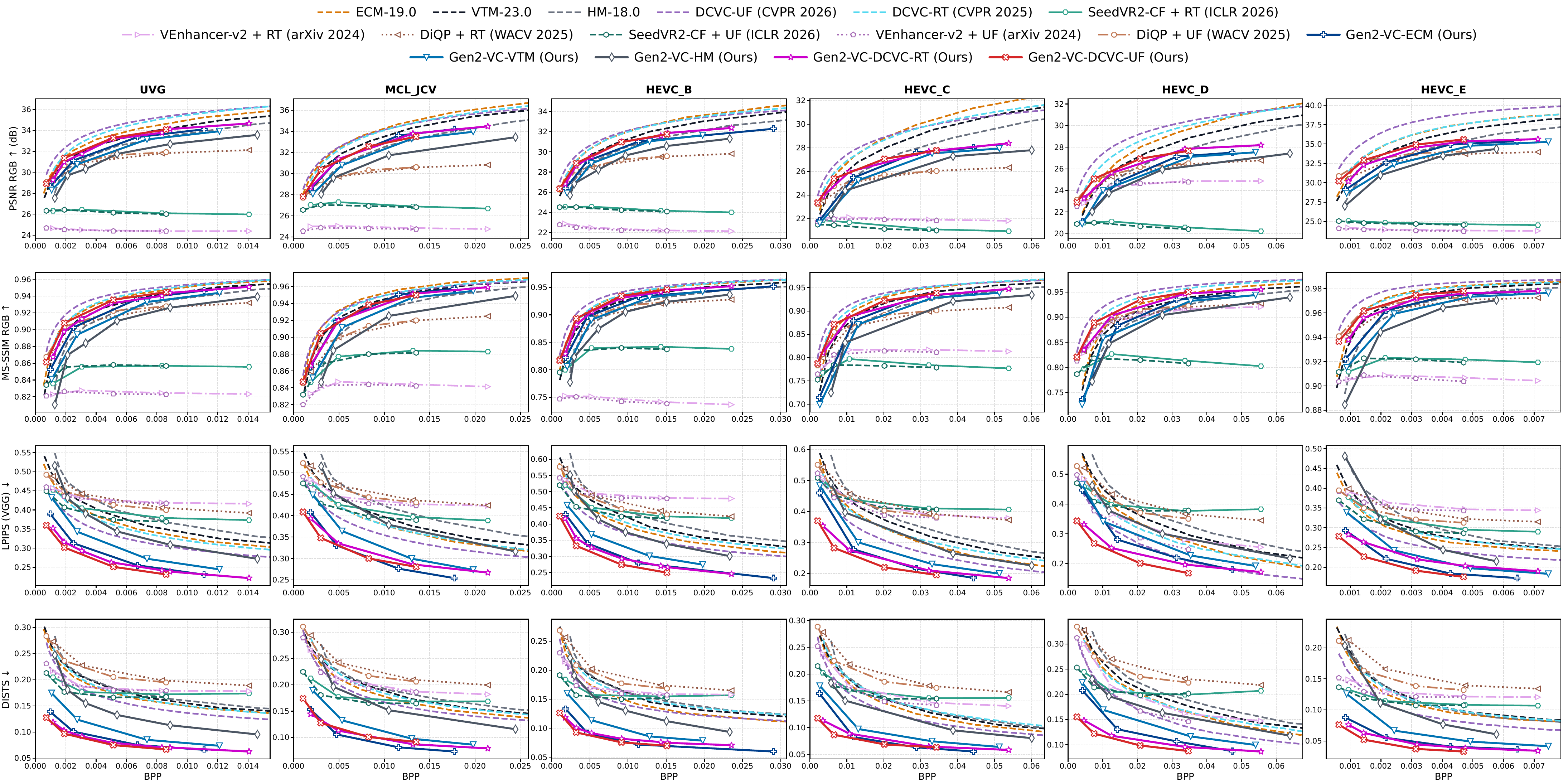}
  \caption{Full-reference comparisons across codec and restoration configurations.}
  \label{fig:rd_codec_distortion}
\end{figure}

Figure~\ref{fig:rd_codec_distortion} shows that adapters trained only on
DCVC-UF improve LPIPS/DISTS on DCVC-RT, ECM, VTM, and HM without retraining.
Given the same codec inputs, Gen2-VC also achieves a better
fidelity--perception balance than the restoration baselines.

\FloatBarrier

\begin{figure}[!htbp]
  \centering
  \includegraphics[width=\linewidth]{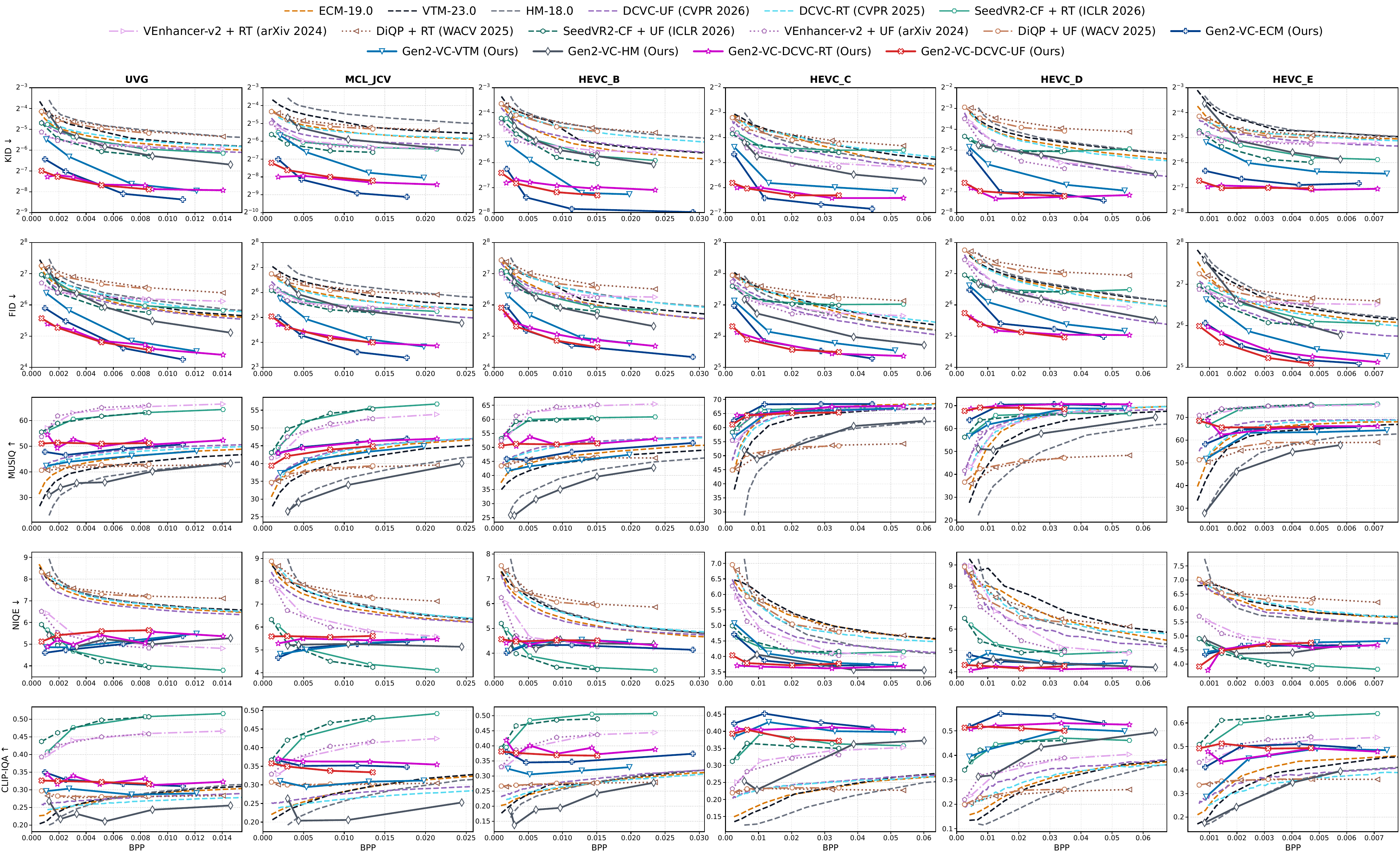}
  \caption{Training-free cross-codec transfer and comparisons with video restoration methods on FID, KID, MUSIQ, NIQE, and CLIP-IQA.}
  \label{fig:rd_codec_perceptual}
\end{figure}

Figure~\ref{fig:rd_codec_perceptual} shows substantially lower FID/KID for
Gen2-VC than the restoration baselines given the same codec inputs.
These gains complement its stronger full-reference performance, although
SeedVR2 and VEnhancer-v2 attain higher MUSIQ or CLIP-IQA at some operating
points.

\FloatBarrier
\subsection{Additional Ablations}
\label{app:additional_ablation}

Table~\ref{tab:ablation_six_dataset} complements the HEVC-E ablations in
Table~\ref{tab:ablation_study} with six-dataset averages.
We select settings for the best joint LPIPS/DISTS performance while retaining
negative six-dataset mean BD-rates in both PSNR and MS-SSIM relative to VTM-23.0
and accounting for LoRA parameter cost.

\begin{figure}[!htbp]
  \centering
  \includegraphics[width=\linewidth]{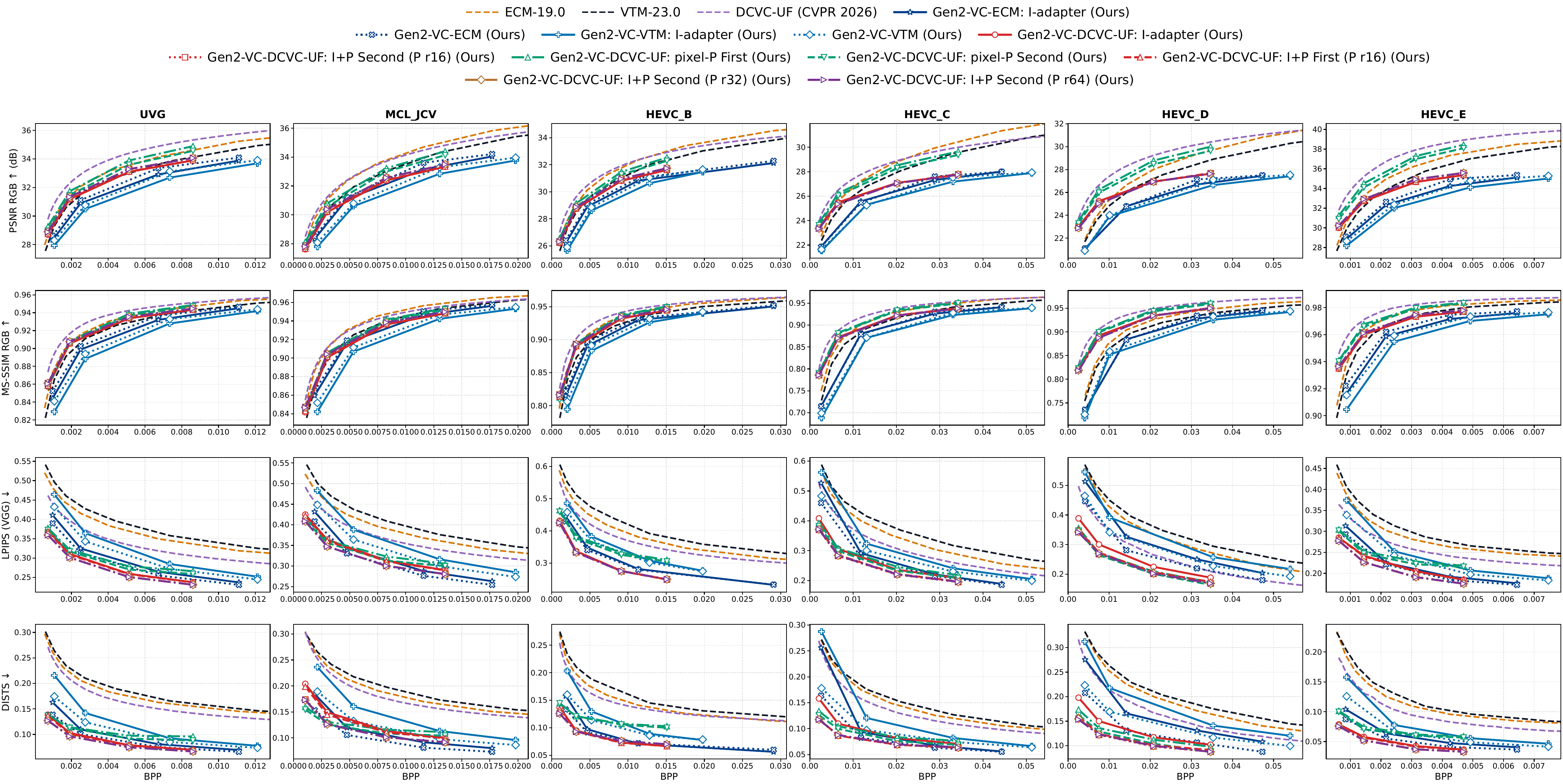}
  \caption{Stage and rank comparisons on six datasets.
  The default \textit{I+P} and \textit{pixel-P} have equal total LoRA parameters.
  At inference, \textit{pixel-P} directly refines native reconstructions without a separate I-adapter.}
  \label{fig:rd_ip_comparison}
\end{figure}
\FloatBarrier

\subsubsection{Stages and Output Selection}
Figure~\ref{fig:rd_ip_comparison} compares \textit{I-only}, which applies the
I-adapter framewise, with \textit{I+P}, which adds recurrent P-adapter refinement
as in Figure~\ref{fig:ip_adaptation}(a).
Following Figure~\ref{fig:ip_adaptation}(b), \textit{First} retains each frame's first
P estimate, whereas \textit{Second} retains its second when available and the first otherwise.
On HEVC-E, Table~\ref{tab:ablation_study}(a) shows that \textit{I+P} \textit{Second}
improves all four BD-rates over \textit{I-only} and \textit{I+P} \textit{First}.
Table~\ref{tab:ablation_six_dataset}(a) confirms this trend in six-dataset averages.
Table~\ref{tab:flolpips_stages} further compares FloLPIPS BD-rates relative to
native DCVC-UF on HEVC-D/C/B, UVG, and MCL-JCV.
\textit{I+P} \textit{Second} achieves the lowest BD-rate on all five datasets,
improving over both \textit{I-only} and \textit{First}.
Together, these results support the P-stage's reconstruction and motion-sensitive
perceptual gains.
Cross-frame comparisons in Appendix~\ref{app:visual_stages} provide complementary
qualitative evidence of consistent detail recovery across sampled frames from
the evaluated 96-frame clips, including after training-free transfer.

To assess adapter separation, \textit{pixel-P} uses one rank-32 adapter for
single-frame spatial training followed by multi-frame temporal training.
At inference, it directly refines native reconstructions without a separate
I-adapter.
Its multi-frame stage shares the P-adapter's training hardware and dataset
mixture, and both pipelines have $11.80$M total LoRA parameters.
With \textit{Second} selection, \textit{I+P} improves mean LPIPS/DISTS BD-rates
from \textit{pixel-P}'s $-81.97\%$/$-90.49\%$ to $-86.65\%$/$-94.24\%$.
With \textit{Second} outputs, \textit{pixel-P} also has worse FloLPIPS BD-rates
than \textit{I+P} on all five datasets in Table~\ref{tab:flolpips_stages},
with the largest gap on HEVC-B.
Under fixed GPU memory, multi-frame training uses lower resolutions than
single-frame spatial training, as summarized in
Table~\ref{tab:appendix_training_config}.
Despite spatial pretraining, \textit{pixel-P} can lose perceptual quality at
higher test resolutions, consistent with this train--test resolution mismatch.
\textit{I+P} instead keeps the independently trained I-adapter fixed, preserving
its higher-resolution spatial refinement in the I-enhanced inputs supplied to
the P-adapter.
This supports the parameter efficiency of separating spatial and temporal adaptation.

\begin{table}[!htbp]
\centering
\caption{Stage and rank ablations on DCVC-UF using four adapter operating points. BD-rate (\%) $\downarrow$ relative to VTM-23.0, averaged equally over UVG, MCL-JCV, HEVC-B, HEVC-C, HEVC-D, and HEVC-E.}
\label{tab:ablation_six_dataset}
\begingroup
\fontsize{9}{10}\selectfont
\setlength{\tabcolsep}{3pt}
\renewcommand{\arraystretch}{1.02}
\begin{tabular*}{\linewidth}{@{\extracolsep{\fill}}llrrrrr@{}}
\toprule
Variant & P output / rank & Total LoRA (M) & PSNR & MS-SSIM & LPIPS & DISTS \\
\midrule
\rowcolor{tablegray}
\multicolumn{7}{c}{\textit{(a) Stages and output selection}} \\
Native DCVC-UF & -- & -- & \textbf{-37.16} & \textbf{-40.04} & -49.41 & -36.39 \\
\textit{I-only} & -- & 5.90 & -9.19 & -24.93 & -83.38 & -90.93 \\
\textit{I+P} & \textit{First} & 11.80 & -8.72 & -24.93 & \underline{-85.03} & \underline{-93.26} \\
\textit{I+P} & \textit{Second} & 11.80 & -12.11 & -28.03 & \textbf{-86.65} & \textbf{-94.24} \\
\textit{pixel-P} (no I) & \textit{First} & 11.80 & \underline{-31.95} & \underline{-34.61} & -80.12 & -89.53 \\
\textit{pixel-P} (no I) & \textit{Second} & 11.80 & -25.26 & -31.80 & -81.97 & -90.49 \\
\midrule
\rowcolor{tablegray}
\multicolumn{7}{c}{\textit{(b) I-adapter rank}} \\
\textit{I-only} & $r_I=8$ & 2.95 & \underline{-9.53} & \underline{-25.22} & -82.64 & \textbf{-91.13} \\
\textit{I-only} (default) & $r_I=16$ & 5.90 & -9.19 & -24.93 & \underline{-83.38} & -90.93 \\
\textit{I-only} & $r_I=32$ & 11.80 & \textbf{-10.72} & \textbf{-26.18} & \textbf{-84.34} & \underline{-90.99} \\
\midrule
\rowcolor{tablegray}
\multicolumn{7}{c}{\textit{(c) P-adapter rank: fixed $r_I=16$, Second outputs}} \\
\textit{I+P} (default) & $r_P=16$ & 11.80 & \textbf{-12.11} & \textbf{-28.03} & \textbf{-86.65} & \textbf{-94.24} \\
\textit{I+P} & $r_P=32$ & 17.69 & \underline{-11.62} & \underline{-27.25} & \underline{-86.61} & \underline{-94.19} \\
\textit{I+P} & $r_P=64$ & 29.49 & -11.59 & -27.20 & -86.60 & -94.18 \\
\bottomrule
\end{tabular*}
\endgroup
\end{table}

\begin{table}[!htbp]
\centering
\caption{FloLPIPS BD-rate (\%) $\downarrow$ relative to native DCVC-UF, using four adapter operating points and 96-frame clips. HEVC-E is omitted because the \textit{I-only} and \textit{I+P} curves have no overlapping quality range with the anchor.}
\label{tab:flolpips_stages}
\begingroup
\fontsize{9}{10}\selectfont
\setlength{\tabcolsep}{4pt}
\renewcommand{\arraystretch}{1.08}
\begin{tabular*}{\linewidth}{@{\extracolsep{\fill}}llrrrrr@{}}
\toprule
Variant & P output & HEVC-D & HEVC-C & HEVC-B & UVG & MCL-JCV \\
\midrule
Native DCVC-UF (Anchor) & -- & 0.00 & 0.00 & 0.00 & 0.00 & 0.00 \\
\textit{I-only} & -- & -54.60 & -54.71 & -76.36 & -83.76 & -63.65 \\
\textit{I+P} & \textit{First} & -66.28 & \underline{-73.83} & \underline{-77.87} & \underline{-84.56} & -67.12 \\
\textit{I+P} & \textit{Second} & \textbf{-68.02} & \textbf{-76.15} & \textbf{-81.71} & \textbf{-85.48} & \textbf{-77.37} \\
\textit{pixel-P} (no I) & \textit{Second} & \underline{-67.09} & -72.31 & -67.38 & -82.24 & \underline{-76.70} \\
\bottomrule
\end{tabular*}
\endgroup
\end{table}

\subsubsection{LoRA Rank}
\label{app:rank_ablation}
Table~\ref{tab:ablation_six_dataset}(b) supports $r_I=16$ as a compromise
between joint LPIPS/DISTS performance and parameter count.
Rank 32 improves both perceptual metrics over rank 16 but doubles its LoRA parameters.
Figure~\ref{fig:rd_ip_comparison} and Table~\ref{tab:ablation_six_dataset}(c) compare P-adapter configurations at ranks 16,
32, and 64, with $r_I=16$, $\alpha_{\mathrm{LoRA}}=2r_P$, and \textit{Second} outputs.
Rank 16 achieves the lowest mean BD-rate on each of the four metrics with
$11.80$M total LoRA parameters, compared with $17.69$M and $29.49$M for ranks
32 and 64, respectively.
With the same hardware, datasets, and training recipe, larger ranks offer
no further mean BD-rate gains, supporting $r_P=16$ as our default.
\FloatBarrier

\FloatBarrier
\clearpage
\subsection{User Study}
\label{app:user_study}

\paragraph{Protocol and stimuli.}
We conducted an anonymous, reference-guided pairwise study with 30 consenting adults (18--34 years), yielding 1,920 formal judgments. 
Participants, mostly university students, were recruited online and volunteered without compensation.
They used personal devices remotely in bright, quiet environments and knew
their responses would be used for experimental analysis.
All completed formal sessions were retained and practice data were excluded.
We evaluated the eight methods listed in Table~\ref{tab:user_study_results} on all 16 HEVC-B/C/D/E sequences (5/4/4/3), using 96-frame clips at native 24--60 fps (1.6--4.0 s).
Gen2-VC-DCVC-UF, DCVC-UF and its SeedVR2 refinement shared the same bitrates. 
Other baselines used the nearest available bitrate at or above ours for each sequence.

\begin{figure}[!htbp]
    \centering
    \includegraphics[width=\linewidth]{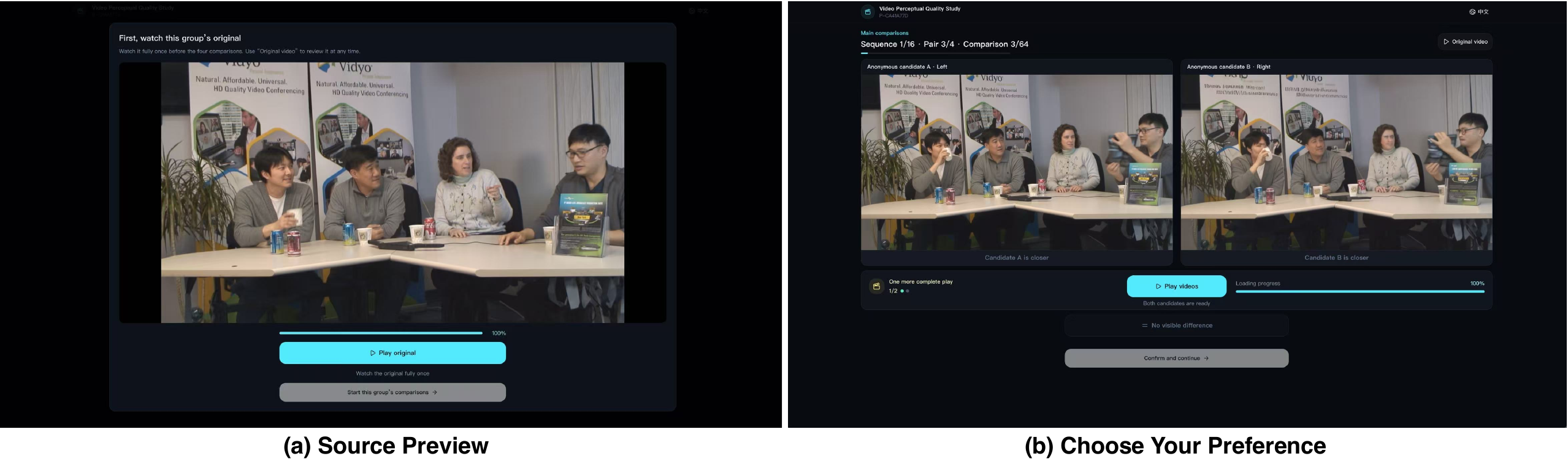}
    \caption{User-study interface. (a) Source preview. (b) Anonymized candidate videos with playback controls and preference choices.}
    \label{fig:user_study_interface}
\end{figure}

After four practice comparisons, participants completed 16 blocks of four comparisons (64 total), with breaks. Each block began with a complete source preview, available for replay throughout. Two anonymized candidates were then played synchronously twice before responses were enabled: left, right, or ``no noticeable visual difference.'' 
Instructions emphasized fidelity to source content and structure, naturalness, artifacts, and temporal stability.
Method identities and bitrates were hidden.
Figure~\ref{fig:user_study_interface} shows the source preview and pairwise comparison interface.

\par\smallskip
\noindent\begin{minipage}{\linewidth}
\centering
\captionof{table}{Preference among eight methods over 16 sequences at selected operating points.}
\label{tab:user_study_results}
\small
\setlength{\tabcolsep}{3pt}
\renewcommand{\arraystretch}{1.08}
\begin{tabular*}{\linewidth}{@{\extracolsep{\fill}}lrrrr@{}}
\toprule
Method & Pref. index $\uparrow$ & 95\% CI & Mean BPP & BPP range \\
\midrule
\textbf{Gen2-VC-DCVC-UF} & \textbf{90.00} & [87.56, 92.38] & 0.00478 & [0.00101, 0.00922] \\
DCVC-UF & 65.17 & [61.75, 68.74] & 0.00478 & [0.00101, 0.00922] \\
ECM-19.0 & 57.18 & [53.54, 60.66] & 0.00635 & [0.00121, 0.01509] \\
SeedVR2 + DCVC-UF & 51.69 & [47.25, 56.28] & 0.00478 & [0.00101, 0.00922] \\
GVC1D & 45.87 & [42.62, 48.96] & 0.00691 & [0.00121, 0.01477] \\
GVC-RT & 42.39 & [39.34, 45.33] & 0.00529 & [0.00102, 0.00939] \\
GLC-Video & 34.68 & [31.15, 38.54] & 0.00877 & [0.00325, 0.01741] \\
YODA & 13.01 & [10.26, 15.89] & 0.00850 & [0.00328, 0.01976] \\
\bottomrule
\end{tabular*}
\par\smallskip
\raggedright\footnotesize
\end{minipage}
\par\smallskip

\paragraph{Pairing and randomization.}
For each sequence, the 28 ($\binom{8}{2}=28$) method pairs  were partitioned into seven units of four disjoint pairs. 
Participants received one unit per sequence, cyclically assigned by formal enrollment using frozen sequence-specific schedules.
Sequence and within-block orders were randomized.
Seeded left/right assignments placed each method eight times on each side per participant. Orientations reversed in successive seven-enrollment cycles.
In the retained cohort, every method pair received 3--5 judgments per sequence.

\paragraph{Analysis.}
We fitted a Davidson model with method strengths $w_a>0$, $w_b>0$, a shared tie parameter $\nu>0$, and a global left-position bias $\delta$. 
For methods $a$ (left) and $b$ (right), the outcome probabilities were
\begin{equation}
 (p_L,p_R,p_T)=
 \frac{\bigl(w_a e^{\delta/2},\,w_b e^{-\delta/2},\,
 \nu\sqrt{w_aw_b}\bigr)}
 {w_a e^{\delta/2}+w_b e^{-\delta/2}+\nu\sqrt{w_aw_b}}.
 \label{eq:user_study_davidson}
\end{equation}
After maximum-likelihood fitting with a small \(L_2\) penalty (\(\lambda=10^{-6}\)), we set $\delta=0$ in prediction and computed the position-neutral preference index
\begin{equation}
 S_a=\frac{100}{7}\sum_{b\ne a}
 \left[P_0(a\succ b)+\tfrac12 P_0(a\sim b)\right].
 \label{eq:user_study_preference_index}
\end{equation}
This model-based index summarizes relative preference, so $90.00$ does not denote a $90\%$ observed win rate.
We obtained 95\% percentile confidence intervals from 10,000 participant-cluster bootstrap refits, retaining each sampled participant's complete set of judgments.

\paragraph{Results.}
Gen2-VC-DCVC-UF achieved the highest preference index (90.00) among the eight
tested methods.
Its mean rate (0.00478 BPP) was joint lowest, matching DCVC-UF and
SeedVR2 + DCVC-UF and remaining below the other five baselines.

\FloatBarrier

\clearpage
\subsection{Computational Complexity}
\label{app:complexity_protocol}
\paragraph{Timing.}
We measure latency with synchronized PyTorch CUDA events after warmup.
Baselines use the mean of ten calls, except GVC-RT, which uses the median.
Latency $t$ in milliseconds gives $1000n/t$ FPS, where $n$ is the number of
new frames per call in Table~\ref{tab:complexity_timing_protocol}.
Timing excludes model loading, disk I/O, and container serialization
and does not capture all host-side overhead.

\paragraph{Profiled calls and bitstream coverage.}
Table~\ref{tab:complexity_timing_protocol} lists new frames per call and
entropy-coding coverage.
S$^2$VC and VoRTeC+ measurements cover P-frames and P-groups, respectively,
excluding initial I-frame/group processing.
Timings without entropy coding are for reference only.
VoRTeC+ additionally excludes AIN color postprocessing.

\par\medskip
\noindent\begin{minipage}{\linewidth}
\centering
\captionof{table}{New frames per call and entropy-coding coverage in timing.}
\label{tab:complexity_timing_protocol}
\begingroup
\small
\setlength{\tabcolsep}{2pt}
\renewcommand{\arraystretch}{1.12}
\begin{tabular*}{\linewidth}{@{\extracolsep{\fill}}lccccccc@{}}
\toprule
\shortstack[l]{Method /\\stage} &
\shortstack{YODA\\GVC-RT\\DCVC-RT\\S$^2$VC} &
DCVC-UF &
VoRTeC+ &
\shortstack{GLC-Video\\GVC1D} &
\shortstack{SeedVR2\\VEnhancer-v2} &
\shortstack{DiQP\\Gen2-VC\\I-adapter} &
\shortstack{Gen2-VC\\P-adapter} \\
\midrule
New frames & 1 & 8 & 8 & 1 & 9 & 1 & 4 \\
Entropy coding & Included & Included & Excluded & Excluded & - & - & - \\
\bottomrule
\end{tabular*}
\endgroup
\end{minipage}
\par\medskip

\paragraph{MACs, memory, and parameters.}
Table~\ref{tab:resource_implementations} summarizes the profiling tools and
accounting rules. MACs are measured separately from latency and memory.

\par\medskip
\noindent\begin{minipage}{\linewidth}
\centering
\captionof{table}{Resource profiling libraries and measurement settings.}
\label{tab:resource_implementations}
\begingroup
\small
\setlength{\tabcolsep}{4pt}
\renewcommand{\arraystretch}{1.12}
\begin{tabular}{@{}>{\raggedright\arraybackslash}p{0.14\linewidth}>{\raggedright\arraybackslash}p{0.36\linewidth}>{\raggedright\arraybackslash}p{\dimexpr0.50\linewidth-4\tabcolsep\relax}@{}}
\toprule
Resource & Implementation & Measurement settings \\
\midrule
MACs &
PyTorch \path{torch.utils.flop_counter.FlopCounterMode} &
Codecs, SeedVR2, and Gen2-VC. Supported FLOPs divided by $2n$ give
MACs per new frame. Codec counts use unfused reference execution. \\
\addlinespace[3pt]
MACs &
Custom PyTorch forward hooks &
DiQP and VEnhancer-v2. Count convolution, transposed-convolution, and
linear modules, then divide by $n$. \\
\addlinespace[3pt]
Peak memory &
PyTorch CUDA \path{reset_peak_memory_stats()},
\path{max_memory_allocated()}, \path{max_memory_reserved()} &
Maximum allocated/reserved GPU memory over measured calls, including
resident models. Gen2-VC uses two separate trials of 24 warmups and
30 measurements, reporting the maximum over 60 measured calls. \\
\addlinespace[3pt]
Parameters &
PyTorch \texttt{numel()} with shared-tensor deduplication &
Total model parameters and trainable adapter parameters.
I/P totals share frozen parameters and should not be summed. \\
\bottomrule
\end{tabular}
\endgroup
\end{minipage}
\par\medskip

\paragraph{Hardware and scope.}
Table~\ref{tab:complexity_full} reports component costs on one A100 80\,GB GPU
for $416\times240$, $832\times480$, $1280\times720$, and $1920\times1080$
inputs, with method-specific padding or resizing.
Restoration methods and Gen2-VC report enhancement-only costs, with I/P
profiled separately and P excluding I. SeedVR2 includes wavelet color correction as described
in Appendix~\ref{app:seedvr2_color_fix}.

\begin{table}[!htbp]
\centering
\caption{Component complexity on one A100 80\,GB GPU. MACs: TMAC/frame. Memory: peak allocated/reserved (A/R), GiB. Parameters: total (trainable), M. S$^2$VC and VoRTeC+ use P-frame/group profiling. GLC-Video, GVC1D, and VoRTeC+ exclude entropy coding. Restoration methods and Gen2-VC report enhancement-only costs.}
\label{tab:complexity_full}
\begingroup
\fontsize{9}{10}\selectfont
\setlength{\tabcolsep}{3.5pt}
\renewcommand{\arraystretch}{0.90}
\begin{tabular*}{\linewidth}{@{\extracolsep{\fill}}llrrrrrr@{}}
\toprule
 & & \multicolumn{2}{c}{FPS $\uparrow$} & \multicolumn{2}{c}{TMAC/frame $\downarrow$} & Memory $\downarrow$ & Params. $\downarrow$ \\
\cmidrule(lr){3-4}\cmidrule(lr){5-6}
Method & Res. & Enc. & Dec. & Enc. & Dec. & A/R & Total (Train.) \\
\midrule
\rowcolor{tablegray}
\multicolumn{8}{c}{\textit{Distortion-oriented video codec}} \\
DCVC-RT & 240p & 223.240 & 233.226 & 0.015 & 0.017 & 0.18/0.29 & 66.33 \\
 & 480p & 222.254 & 233.857 & 0.060 & 0.069 & 0.23/0.30 &  \\
 & 720p & 212.467 & 173.954 & 0.138 & 0.160 & 0.31/0.42 &  \\
 & 1080p & 130.670 & 119.125 & 0.312 & 0.362 & 0.49/0.74 &  \\
\addlinespace[1pt]
DCVC-UF & 240p & 2848.574 & 1998.211 & 0.013 & 0.015 & 0.37/0.40 & 162.72 \\
 & 480p & 1350.545 & 1014.879 & 0.054 & 0.058 & 0.54/0.62 &  \\
 & 720p & 666.658 & 537.972 & 0.124 & 0.134 & 0.85/1.04 &  \\
 & 1080p & 330.183 & 274.652 & 0.282 & 0.305 & 1.46/1.88 &  \\
\midrule
\rowcolor{tablegray}
\multicolumn{8}{c}{\textit{VAE/tokenizer-based video codec (analysis/synthesis only, excluding entropy coding)}} \\
GLC-Video & 240p & 53.393 & 55.358 & 0.139 & 0.223 & 0.75/1.29 & 289.33 \\
 & 480p & 32.391 & 21.277 & 0.519 & 0.830 & 1.22/2.01 &  \\
 & 720p & 16.140 & 9.583 & 1.196 & 1.913 & 2.05/3.65 &  \\
 & 1080p & 8.059 & 4.719 & 2.546 & 4.071 & 3.69/7.14 &  \\
\addlinespace[1pt]
GVC1D & 240p & 11.433 & 32.568 & 0.202 & 0.434 & 5.91/6.08 & 1800.93 \\
 & 480p & 10.663 & 18.904 & 0.856 & 1.833 & 8.38/8.44 &  \\
 & 720p & 5.385 & 4.151 & 1.712 & 3.653 & 14.87/14.99 &  \\
 & 1080p & 3.086 & 1.941 & 5.588 & 11.786 & 69.81/70.10 &  \\
\midrule
\rowcolor{tablegray}
\multicolumn{8}{c}{\textit{VAE/tokenizer-based video codec (including entropy coding)}} \\
GVC-RT & 240p & 224.368 & 106.927 & 0.017 & 0.039 & 0.32/0.43 & 135.17 \\
 & 480p & 208.311 & 114.131 & 0.064 & 0.145 & 0.36/0.49 &  \\
 & 720p & 189.128 & 95.849 & 0.147 & 0.334 & 0.46/0.58 &  \\
 & 1080p & 127.115 & 49.967 & 0.312 & 0.710 & 0.64/0.86 &  \\
\midrule
\rowcolor{tablegray}
\multicolumn{8}{c}{\textit{Diffusion-based video codec}} \\
YODA & 240p & 3.958 & 5.011 & 2.419 & 1.601 & 9.85/10.49 & 2055.13 \\
 & 480p & 2.860 & 3.920 & 4.469 & 2.951 & 11.64/12.93 &  \\
 & 720p & 1.536 & 2.292 & 10.279 & 6.776 & 16.69/21.37 &  \\
 & 1080p & 0.780 & 1.230 & 21.813 & 14.370 & 26.73/39.32 &  \\
\addlinespace[1pt]
S$^2$VC & 240p & 4.857 & 5.600 & 0.918 & 0.859 & 7.32/8.09 & 1842.29 \\
 & 480p & 2.148 & 2.483 & 3.532 & 3.310 & 8.25/9.41 &  \\
 & 720p & 0.987 & 1.113 & 8.650 & 8.138 & 9.94/12.50 &  \\
 & 1080p & 0.404 & 0.439 & 20.490 & 19.403 & 13.31/18.61 &  \\
\addlinespace[1pt]
VoRTeC+ & 240p & 9.341 & 6.312 & 0.344 & 0.893 & 9.24/10.01 & 1802.72 \\
 & 480p & 0.977 & 0.482 & 1.288 & 3.551 & 14.62/17.54 &  \\
 & 720p & 0.476 & 0.240 & 3.011 & 9.271 & 24.18/29.76 &  \\
 & 1080p & 0.188 & 0.097 & 6.561 & 24.290 & 43.20/56.00 &  \\
\midrule
\rowcolor{tablegray}
\multicolumn{8}{c}{\textit{Video restoration methods}} \\
SeedVR2 & 240p & -- & 0.368 & -- & 2.014 & 36.36/36.74 & 3642.12 \\
 & 480p & -- & 0.344 & -- & 8.028 & 39.94/40.55 &  \\
 & 720p & -- & 0.258 & -- & 18.586 & 46.87/49.11 &  \\
 & 1080p & -- & 0.194 & -- & 42.515 & 68.37/78.09 &  \\
\addlinespace[1pt]
DiQP & 240p & -- & 3.050 & -- & 0.589 & 1.48/2.15 & 79.37 \\
 & 480p & -- & 1.695 & -- & 1.178 & 1.48/2.15 &  \\
 & 720p & -- & 0.527 & -- & 3.534 & 1.48/2.15 &  \\
 & 1080p & -- & 0.261 & -- & 7.067 & 1.48/2.15 &  \\
\addlinespace[1pt]
VEnhancer-v2 & 240p & -- & 0.072 & -- & 145.383 & 19.55/78.62 & 2496.59 \\
 & 480p & -- & 0.072 & -- & 145.383 & 19.58/78.62 &  \\
 & 720p & -- & 0.071 & -- & 145.383 & 19.63/78.62 &  \\
 & 1080p & -- & 0.093 & -- & 216.563 & 23.81/78.62 &  \\
\midrule
\rowcolor{tablegray}
\multicolumn{8}{c}{\textit{Gen2-VC variants}} \\
Gen2-VC I & 240p & -- & 16.659 & -- & 2.071 & 3.27/5.42 & 1551.79 (5.90) \\
 & 480p & -- & 5.234 & -- & 8.205 & 3.29/12.56 &  \\
 & 720p & -- & 2.274 & -- & 20.268 & 3.34/24.20 &  \\
 & 1080p & -- & 0.887 & -- & 53.519 & 10.83/27.21 &  \\
\addlinespace[1pt]
Gen2-VC P & 240p & -- & 12.807 & -- & 2.430 & 5.21/7.37 & 1551.79 (5.90) \\
 & 480p & -- & 3.553 & -- & 10.396 & 11.22/19.35 &  \\
 & 720p & -- & 1.477 & -- & 27.122 & 21.67/39.08 &  \\
 & 1080p & -- & 0.557 & -- & 77.546 & 45.25/69.17 &  \\
\bottomrule
\end{tabular*}
\par\vspace{3pt}
\endgroup
\end{table}

\FloatBarrier
\par\medskip
\noindent\begin{minipage}{\linewidth}
\paragraph{Complete Decoding Pipeline.}
Table~\ref{tab:complexity_pipeline} includes DCVC-UF decoding followed by I/P enhancement.
On one GPU, the adapters run serially and share one resident Wan backbone.
In the two-GPU pipeline, GPU A runs codec decoding and the I-adapter, while GPU B
runs the P-adapter, with one resident Wan backbone per GPU.
This split is enabled by keeping enhancement outside the codec's reference chain.
In diffusion-based codecs that couple refinement with latent reference updates,
temporal dependencies constrain such pipelining.
\par\smallskip
\centering
\captionof{table}{Complete decoding costs of Gen2-VC-DCVC-UF. FPS is measured, memory is estimated, and MACs per frame are identical for both configurations.}
\label{tab:complexity_pipeline}
\begingroup
\fontsize{9}{10}\selectfont
\setlength{\tabcolsep}{3.5pt}
\renewcommand{\arraystretch}{1.12}
\begin{tabular*}{\linewidth}{@{\extracolsep{\fill}}lrrrrr@{}}
\toprule
 & \multicolumn{2}{c}{FPS $\uparrow$} & TMAC/frame $\downarrow$ & \multicolumn{2}{c}{Estimated memory (GiB) $\downarrow$} \\
\cmidrule(lr){2-3}\cmidrule(lr){5-6}
Res. & Single GPU & Two GPUs & Both & Single GPU & Two GPUs: A / B \\
\midrule
240p & 7.215 & 12.807 & 4.516 & 6.02 & 4.60 / 5.70 \\
480p & 2.112 & 3.553 & 18.660 & 12.43 & 6.69 / 12.12 \\
720p & 0.894 & 1.477 & 47.525 & 23.61 & 10.32 / 23.30 \\
1080p & 0.342 & 0.557 & 131.370 & 45.76 & 14.19 / 45.45 \\
\bottomrule
\end{tabular*}
\endgroup
\end{minipage}
\par\medskip

\paragraph{Cost interpretation.}
Counting shared weights once gives approximately $1.72$B total parameters,
below the totals of all three diffusion codecs in Table~\ref{tab:complexity_full}.
Only $11.80$M LoRA parameters are trained.
Native DCVC-UF/RT decoding contributes little computation relative to I/P enhancement.
Complete single-GPU throughput exceeds profiled VoRTeC+ decoding and all three
restoration stages across resolutions, but trails S$^2$VC/YODA at 480p--1080p.
Two GPUs improve throughput without reducing MACs or window buffering.

\paragraph{Buffering and output latency.}
Following Figure~\ref{fig:ip_adaptation}(b), frames retained for a \textit{Second}
estimate wait for the next overlapping call and four new inputs, adding
buffering beyond native low-delay decoding.
The reported FPS measures throughput, not per-frame output latency.

\FloatBarrier

\subsection{Per-Dataset BD-Rates}
\label{app:per_dataset_bd_rate}
Tables~\ref{tab:bd_rate_per_dataset}, \ref{tab:bd_rate_per_dataset_hevc_b_c},
and~\ref{tab:bd_rate_per_dataset_hevc_d_e} report PSNR, MS-SSIM, LPIPS,
and DISTS BD-rates for the 22 methods in Table~\ref{tab:bd_rate_summary}.
The default uses $r_I=r_P=16$ and retains \textit{Second} outputs.
Gen2-VC-DCVC-UF combines strong LPIPS/DISTS performance with substantially
better PSNR/MS-SSIM than the evaluated generative codecs across all six datasets.
Training-free transfer also improves LPIPS and DISTS BD-rates over native
DCVC-RT, ECM, VTM, and HM on all six datasets, demonstrating consistent
perceptual gains across codec families.

\paragraph{Aggregation.}
Table~\ref{tab:bd_rate_summary} averages each metric's BD-rates equally across
all six datasets using unrounded values.
We report ``--'' for non-overlapping quality ranges or a six-dataset mean
with any missing dataset result.

\begin{table}[!htbp]
\centering
\caption{BD-rate (\%) $\downarrow$ relative to VTM-23.0 on UVG and MCL-JCV.}
\label{tab:bd_rate_per_dataset}
\begingroup
\fontsize{8}{9}\selectfont
\setlength{\tabcolsep}{1.35pt}
\renewcommand{\arraystretch}{0.96}
\begin{tabular*}{\linewidth}{@{\extracolsep{\fill}}lrrrr@{\hspace{6pt}}rrrr@{}}
\toprule
Method & \multicolumn{4}{c}{UVG} & \multicolumn{4}{c}{MCL-JCV} \\
\cmidrule(lr){2-5}\cmidrule(lr){6-9}
 & PSNR & MS-SSIM & LPIPS & DISTS & PSNR & MS-SSIM & LPIPS & DISTS \\
\midrule
\rowcolor{tablegray}
\multicolumn{9}{c}{\textit{Distortion-oriented video codec}} \\
HM-18.0 & $39.78$ & $46.17$ & $30.67$ & $13.05$ & $44.54$ & $53.35$ & $40.10$ & $16.98$ \\
VTM-23.0 (Anchor) & $0.00$ & $0.00$ & $0.00$ & $0.00$ & $0.00$ & $0.00$ & $0.00$ & $0.00$ \\
ECM-19.0 & $-20.13$ & $-19.09$ & $-18.15$ & $-14.64$ & $\mathbf{-22.37}$ & $\mathbf{-22.53}$ & $-21.70$ & $-16.56$ \\
DCVC-RT & \underline{$-30.52$} & $-21.19$ & $-25.92$ & $-6.23$ & $-13.74$ & $-8.64$ & $-18.85$ & $3.73$ \\
DCVC-UF & $\mathbf{-41.55}$ & $\mathbf{-37.28}$ & $-51.85$ & $-34.67$ & \underline{$-20.98$} & \underline{$-19.43$} & $-46.03$ & $-26.70$ \\
\midrule
\rowcolor{tablegray}
\multicolumn{9}{c}{\textit{VAE/tokenizer-based video codec}} \\
GLC-Video & $858.40$ & $570.84$ & $-54.19$ & $-93.04$ & $504.62$ & $291.25$ & $-71.92$ & $-92.94$ \\
GVC1D & $413.50$ & $322.89$ & $-80.02$ & $-94.49$ & $219.05$ & $159.68$ & $\mathbf{-89.27}$ & $\mathbf{-96.42}$ \\
GVC-RT & $1127.30$ & $175.55$ & $-83.81$ & $-94.27$ & $846.23$ & $132.34$ & \underline{$-86.33$} & $-92.37$ \\
\midrule
\rowcolor{tablegray}
\multicolumn{9}{c}{\textit{Diffusion-based video codec}} \\
YODA & $1219.48$ & $1139.05$ & $-56.46$ & $\mathbf{-97.64}$ & $712.53$ & $634.29$ & $-76.76$ & \underline{$-95.76$} \\
S$^2$VC & $384.01$ & $167.87$ & $-62.92$ & $-83.40$ & $257.01$ & $122.46$ & $-78.05$ & $-88.40$ \\
VoRTeC+ & $547.74$ & $300.17$ & $-71.34$ & $-91.12$ & $412.10$ & $209.04$ & $-82.13$ & $-92.41$ \\
\midrule
\rowcolor{tablegray}
\multicolumn{9}{c}{\textit{Video restoration methods}} \\
\shortstack[l]{SeedVR2 +\\DCVC-RT} & -- & $108.38$ & $-19.29$ & $-56.59$ & -- & $93.89$ & $-19.25$ & $-63.54$ \\
\shortstack[l]{VEnhancer-v2 +\\DCVC-RT} & -- & $917.69$ & $4.18$ & $-44.42$ & -- & $953.66$ & $-3.31$ & $-14.19$ \\
\shortstack[l]{DiQP +\\DCVC-RT} & $63.86$ & $19.51$ & $49.66$ & $63.10$ & $102.89$ & $57.52$ & $61.19$ & $61.04$ \\
\shortstack[l]{SeedVR2 +\\DCVC-UF} & -- & $42.65$ & $-48.00$ & $-68.08$ & -- & $40.23$ & $-49.46$ & $-76.18$ \\
\shortstack[l]{VEnhancer-v2 +\\DCVC-UF} & -- & $522.58$ & $-23.22$ & $-56.86$ & -- & $754.43$ & $-36.55$ & $-31.93$ \\
\shortstack[l]{DiQP +\\DCVC-UF} & $11.94$ & $-15.34$ & $2.14$ & $18.70$ & $40.60$ & $13.64$ & $7.54$ & $19.08$ \\
\midrule
\rowcolor{tablegray}
\multicolumn{9}{c}{\textit{Gen2-VC variants}} \\
Gen2-VC-HM & $96.18$ & $102.12$ & $-24.58$ & $-60.39$ & $106.45$ & $96.69$ & $-2.55$ & $-44.22$ \\
Gen2-VC-VTM & $45.48$ & $43.83$ & $-70.56$ & $-89.71$ & $48.56$ & $36.44$ & $-64.82$ & $-84.92$ \\
Gen2-VC-ECM & $16.85$ & $9.53$ & $-82.89$ & $-95.69$ & $14.78$ & $2.57$ & $-80.83$ & $-94.01$ \\
\shortstack[l]{Gen2-VC-\\DCVC-RT} & $2.49$ & $-0.29$ & \underline{$-85.83$} & $-96.49$ & $14.28$ & $5.22$ & $-78.42$ & $-92.86$ \\
\shortstack[l]{Gen2-VC-\\DCVC-UF} & $-16.51$ & \underline{$-24.65$} & $\mathbf{-89.33}$ & \underline{$-97.24$} & $1.09$ & $-11.57$ & $-83.97$ & $-92.53$ \\
\bottomrule
\end{tabular*}
\endgroup
\end{table}

\begin{table}[!htbp]
\centering
\caption{BD-rate (\%) $\downarrow$ relative to VTM-23.0 on HEVC-B and HEVC-C.}
\label{tab:bd_rate_per_dataset_hevc_b_c}
\begingroup
\fontsize{8}{9}\selectfont
\setlength{\tabcolsep}{1.35pt}
\renewcommand{\arraystretch}{0.96}
\begin{tabular*}{\linewidth}{@{\extracolsep{\fill}}lrrrr@{\hspace{6pt}}rrrr@{}}
\toprule
Method & \multicolumn{4}{c}{HEVC-B} & \multicolumn{4}{c}{HEVC-C} \\
\cmidrule(lr){2-5}\cmidrule(lr){6-9}
 & PSNR & MS-SSIM & LPIPS & DISTS & PSNR & MS-SSIM & LPIPS & DISTS \\
\midrule
\rowcolor{tablegray}
\multicolumn{9}{c}{\textit{Distortion-oriented video codec}} \\
HM-18.0 & $36.90$ & $45.90$ & $26.20$ & $11.81$ & $36.35$ & $45.26$ & $25.71$ & $11.04$ \\
VTM-23.0 (Anchor) & $0.00$ & $0.00$ & $0.00$ & $0.00$ & $0.00$ & $0.00$ & $0.00$ & $0.00$ \\
ECM-19.0 & \underline{$-21.06$} & $-20.53$ & $-22.09$ & $-21.36$ & \underline{$-22.39$} & $-20.34$ & $-21.49$ & $-12.70$ \\
DCVC-RT & $-18.94$ & $-19.05$ & $-12.54$ & $3.12$ & $-14.72$ & $-29.54$ & $-16.72$ & $10.38$ \\
DCVC-UF & $\mathbf{-28.37}$ & $\mathbf{-37.21}$ & $-44.59$ & $-36.60$ & $\mathbf{-30.75}$ & $\mathbf{-43.61}$ & $-46.32$ & $-31.59$ \\
\midrule
\rowcolor{tablegray}
\multicolumn{9}{c}{\textit{VAE/tokenizer-based video codec}} \\
GLC-Video & $796.76$ & $410.41$ & $-62.19$ & $-91.10$ & $625.07$ & $233.62$ & $-56.91$ & $-70.48$ \\
GVC1D & $491.77$ & $246.58$ & \underline{$-83.82$} & $-95.97$ & $379.48$ & $155.40$ & $-76.34$ & $-85.26$ \\
GVC-RT & $713.26$ & $124.20$ & $-81.45$ & $-92.69$ & $293.91$ & $66.85$ & $-76.60$ & $-87.28$ \\
\midrule
\rowcolor{tablegray}
\multicolumn{9}{c}{\textit{Diffusion-based video codec}} \\
YODA & $1069.66$ & $761.52$ & $-65.77$ & $\mathbf{-98.17}$ & $782.05$ & $375.00$ & $-55.19$ & $-86.77$ \\
S$^2$VC & $368.63$ & $161.10$ & $-69.98$ & $-85.76$ & $236.61$ & $79.93$ & $-60.74$ & $-64.10$ \\
VoRTeC+ & $477.64$ & $170.45$ & $-81.51$ & $-94.67$ & $300.51$ & $85.45$ & \underline{$-79.00$} & \underline{$-87.79$} \\
\midrule
\rowcolor{tablegray}
\multicolumn{9}{c}{\textit{Video restoration methods}} \\
\shortstack[l]{SeedVR2 +\\DCVC-RT} & -- & $624.27$ & $-27.38$ & $-66.17$ & -- & $499.08$ & $-5.55$ & $-28.96$ \\
\shortstack[l]{VEnhancer-v2 +\\DCVC-RT} & -- & -- & $7.77$ & $-40.98$ & -- & $492.70$ & $1.34$ & $-15.52$ \\
\shortstack[l]{DiQP +\\DCVC-RT} & $63.22$ & $12.07$ & $29.82$ & $41.13$ & $59.80$ & $14.62$ & $31.05$ & $78.71$ \\
\shortstack[l]{SeedVR2 +\\DCVC-UF} & -- & $12.83$ & $-53.78$ & $-74.96$ & -- & $399.81$ & $-31.12$ & $-45.95$ \\
\shortstack[l]{VEnhancer-v2 +\\DCVC-UF} & -- & -- & $-34.37$ & $-56.92$ & -- & $-4.19$ & $-32.75$ & $-39.88$ \\
\shortstack[l]{DiQP +\\DCVC-UF} & $17.18$ & $-19.93$ & $-12.37$ & $-2.19$ & $8.41$ & $-20.02$ & $-15.07$ & $24.16$ \\
\midrule
\rowcolor{tablegray}
\multicolumn{9}{c}{\textit{Gen2-VC variants}} \\
Gen2-VC-HM & $95.87$ & $81.20$ & $-32.71$ & $-51.25$ & $98.72$ & $70.36$ & $-32.06$ & $-49.65$ \\
Gen2-VC-VTM & $46.34$ & $24.10$ & $-71.04$ & $-86.31$ & $70.41$ & $28.38$ & $-64.12$ & $-79.73$ \\
Gen2-VC-ECM & $25.32$ & $1.68$ & $-82.65$ & $-94.91$ & $42.61$ & $6.43$ & $-73.62$ & $-86.92$ \\
\shortstack[l]{Gen2-VC-\\DCVC-RT} & $5.95$ & $-9.28$ & $-83.80$ & $-95.23$ & $20.51$ & $-12.71$ & $-76.47$ & $-85.98$ \\
\shortstack[l]{Gen2-VC-\\DCVC-UF} & $-11.09$ & \underline{$-30.02$} & $\mathbf{-88.83}$ & \underline{$-96.66$} & $-7.86$ & \underline{$-32.91$} & $\mathbf{-84.54}$ & $\mathbf{-91.93}$ \\
\bottomrule
\end{tabular*}
\endgroup
\end{table}

\begin{table}[!htbp]
\centering
\caption{BD-rate (\%) $\downarrow$ relative to VTM-23.0 on HEVC-D and HEVC-E.}
\label{tab:bd_rate_per_dataset_hevc_d_e}
\begingroup
\fontsize{8}{9}\selectfont
\setlength{\tabcolsep}{1.35pt}
\renewcommand{\arraystretch}{0.96}
\begin{tabular*}{\linewidth}{@{\extracolsep{\fill}}lrrrr@{\hspace{6pt}}rrrr@{}}
\toprule
Method & \multicolumn{4}{c}{HEVC-D} & \multicolumn{4}{c}{HEVC-E} \\
\cmidrule(lr){2-5}\cmidrule(lr){6-9}
 & PSNR & MS-SSIM & LPIPS & DISTS & PSNR & MS-SSIM & LPIPS & DISTS \\
\midrule
\rowcolor{tablegray}
\multicolumn{9}{c}{\textit{Distortion-oriented video codec}} \\
HM-18.0 & $29.43$ & $31.60$ & $27.38$ & $20.64$ & $42.90$ & $41.29$ & $25.45$ & $-4.23$ \\
VTM-23.0 (Anchor) & $0.00$ & $0.00$ & $0.00$ & $0.00$ & $0.00$ & $0.00$ & $0.00$ & $0.00$ \\
ECM-19.0 & $-21.14$ & $-14.77$ & $-18.64$ & $-14.97$ & $-18.62$ & $-17.01$ & $-16.56$ & $-11.43$ \\
DCVC-RT & \underline{$-28.72$} & $-40.29$ & $-26.12$ & $-12.40$ & \underline{$-21.84$} & $-17.40$ & $-10.37$ & $-5.25$ \\
DCVC-UF & $\mathbf{-48.44}$ & $\mathbf{-56.59}$ & $-54.62$ & $-45.96$ & $\mathbf{-52.88}$ & $\mathbf{-46.08}$ & $-53.05$ & $-42.83$ \\
\midrule
\rowcolor{tablegray}
\multicolumn{9}{c}{\textit{VAE/tokenizer-based video codec}} \\
GLC-Video & -- & $200.56$ & $-46.46$ & $-67.34$ & -- & $483.60$ & $-45.93$ & $-58.32$ \\
GVC1D & $287.09$ & $103.34$ & $-74.94$ & $-84.42$ & $629.33$ & $211.95$ & $-87.09$ & $-90.28$ \\
GVC-RT & $160.30$ & $29.58$ & $-74.28$ & $-85.09$ & $1756.91$ & $210.79$ & \underline{$-89.46$} & $-84.59$ \\
\midrule
\rowcolor{tablegray}
\multicolumn{9}{c}{\textit{Diffusion-based video codec}} \\
YODA & $873.61$ & $499.20$ & $-23.20$ & $-72.45$ & $1248.55$ & $812.81$ & $-65.49$ & $-93.05$ \\
S$^2$VC & $160.73$ & $39.98$ & $-56.78$ & $-63.01$ & $439.50$ & $186.83$ & $-63.31$ & $-80.82$ \\
VoRTeC+ & $170.18$ & $39.12$ & \underline{$-79.32$} & \underline{$-86.49$} & $547.79$ & $285.36$ & $-87.47$ & $-92.27$ \\
\midrule
\rowcolor{tablegray}
\multicolumn{9}{c}{\textit{Video restoration methods}} \\
\shortstack[l]{SeedVR2 +\\DCVC-RT} & -- & $363.19$ & $-24.59$ & $57.76$ & -- & $476.80$ & $-0.73$ & $-41.45$ \\
\shortstack[l]{VEnhancer-v2 +\\DCVC-RT} & $11.62$ & $-18.04$ & $-33.04$ & $-30.38$ & -- & $351.98$ & $52.50$ & $-22.76$ \\
\shortstack[l]{DiQP +\\DCVC-RT} & $12.15$ & $-14.60$ & $15.05$ & $46.13$ & $30.33$ & $8.86$ & $40.90$ & $53.02$ \\
\shortstack[l]{SeedVR2 +\\DCVC-UF} & -- & $204.27$ & $-49.01$ & $-63.97$ & -- & $288.39$ & $-35.68$ & $-50.79$ \\
\shortstack[l]{VEnhancer-v2 +\\DCVC-UF} & $-18.90$ & $-42.93$ & $-55.48$ & $-52.69$ & -- & $367.31$ & $1.85$ & $-50.79$ \\
\shortstack[l]{DiQP +\\DCVC-UF} & $-27.48$ & $-43.05$ & $-29.00$ & $-7.45$ & $-17.02$ & \underline{$-27.74$} & $-8.23$ & $-0.03$ \\
\midrule
\rowcolor{tablegray}
\multicolumn{9}{c}{\textit{Gen2-VC variants}} \\
Gen2-VC-HM & $82.34$ & $50.06$ & $-23.76$ & $-42.04$ & $84.43$ & $73.17$ & $-11.92$ & $-40.83$ \\
Gen2-VC-VTM & $36.86$ & $8.90$ & $-53.06$ & $-70.03$ & $56.97$ & $35.91$ & $-74.40$ & $-83.88$ \\
Gen2-VC-ECM & $40.33$ & $9.59$ & $-61.67$ & $-79.04$ & $31.92$ & $12.71$ & $-88.32$ & $-94.45$ \\
\shortstack[l]{Gen2-VC-\\DCVC-RT} & $-0.85$ & $-28.53$ & $-72.61$ & $-83.95$ & $9.95$ & $1.89$ & $-82.24$ & \underline{$-96.05$} \\
\shortstack[l]{Gen2-VC-\\DCVC-UF} & $-25.42$ & \underline{$-46.54$} & $\mathbf{-81.20}$ & $\mathbf{-89.63}$ & $-12.89$ & $-22.51$ & $\mathbf{-92.06}$ & $\mathbf{-97.47}$ \\
\bottomrule
\end{tabular*}
\endgroup
\end{table}

\FloatBarrier

\clearpage
\subsection{Additional Visual Comparisons}
\label{app:additional_visual}

\subsubsection{Codec Comparisons}
\label{app:visual_codec}
\begin{figure}[!htbp]
  \centering
  \includegraphics[width=\linewidth]{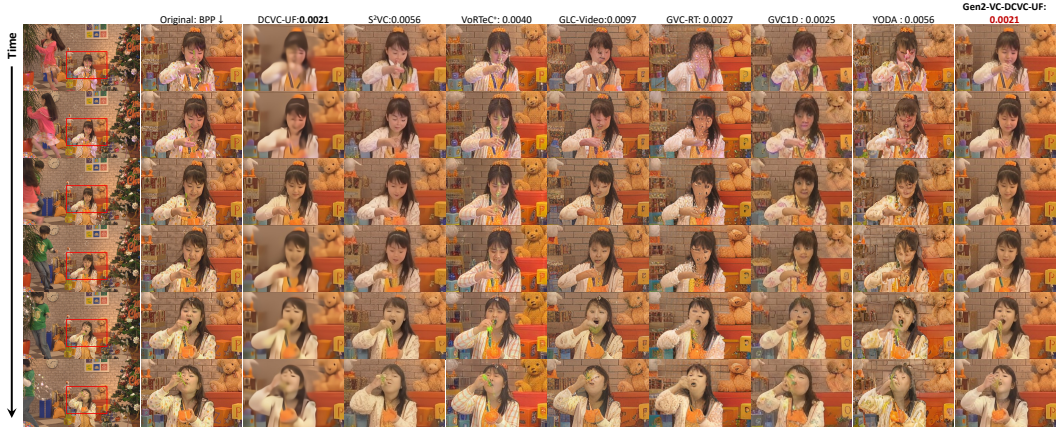}
  \caption{Low-bitrate codec comparison across sampled frames, example~1.}
  \label{fig:app_visual_codec_1}
\end{figure}

\begin{figure}[!htbp]
  \centering
  \includegraphics[width=\linewidth]{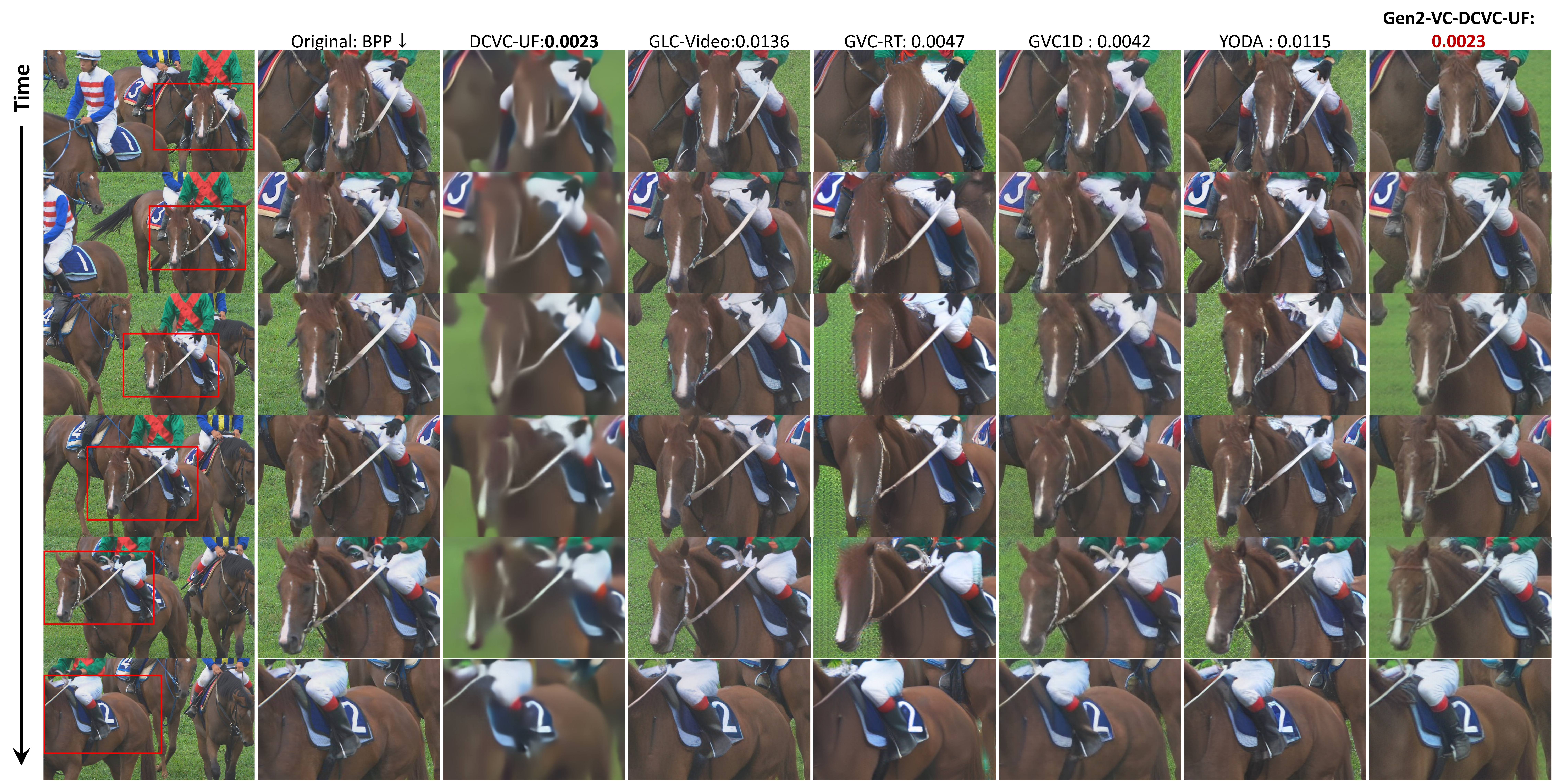}
  \caption{Low-bitrate codec comparison across sampled frames, example~2.}
  \label{fig:app_visual_codec_2}
\end{figure}

\begin{figure}[!htbp]
  \centering
  \includegraphics[width=\linewidth]{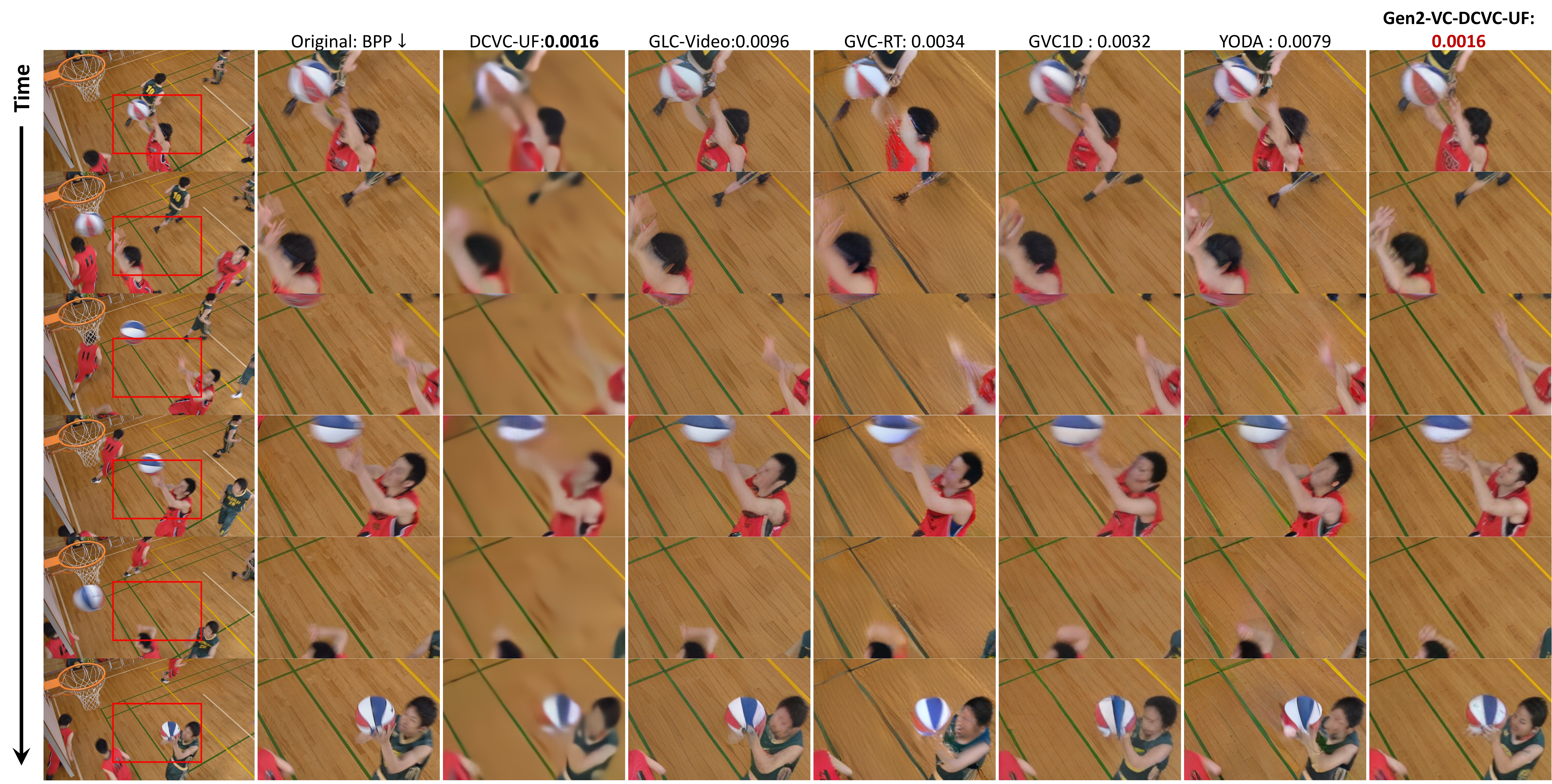}
  \caption{Low-bitrate codec comparison across sampled frames, example~3.}
  \label{fig:app_visual_codec_3}
\end{figure}

\begin{figure}[!htbp]
  \centering
  \includegraphics[width=\linewidth]{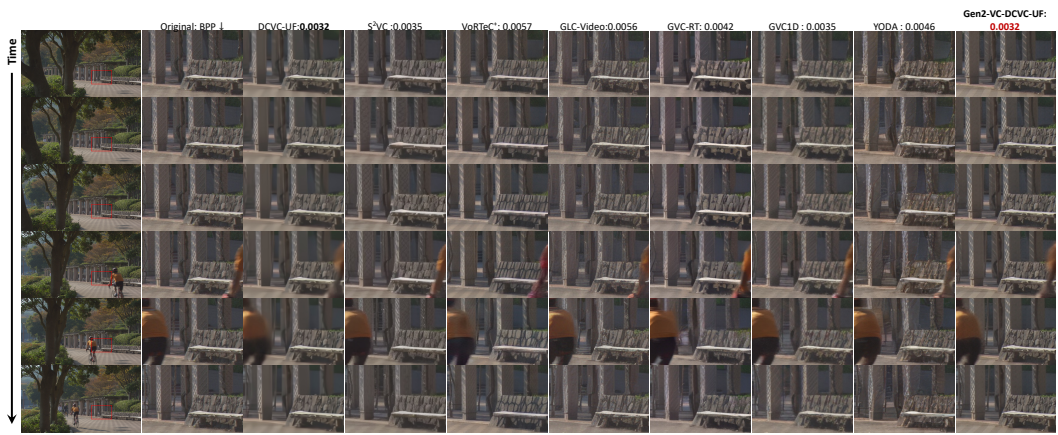}
  \caption{Low-bitrate codec comparison across sampled frames, example~4.}
  \label{fig:app_visual_codec_4}
\end{figure}

\begin{figure}[!htbp]
  \centering
  \includegraphics[width=\linewidth]{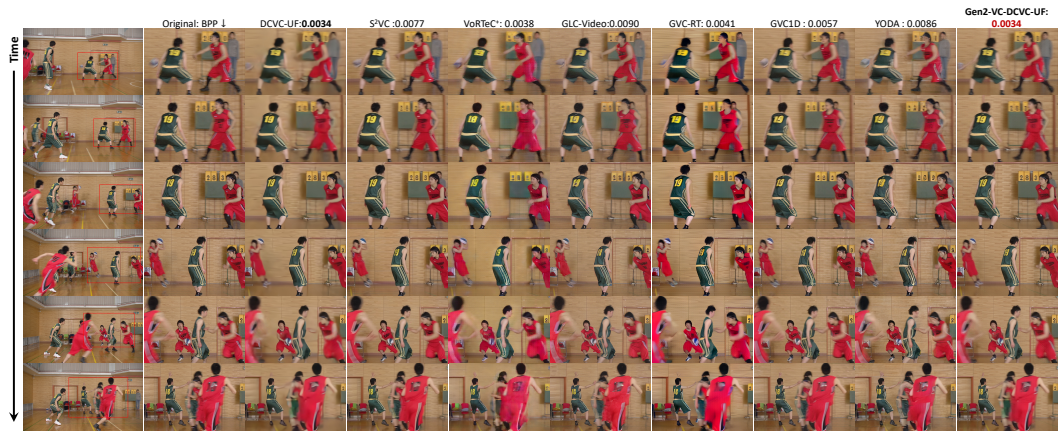}
  \caption{Low-bitrate codec comparison across sampled frames, example~5.}
  \label{fig:app_visual_codec_5}
\end{figure}

\begin{figure}[!htbp]
  \centering
  \includegraphics[width=\linewidth]{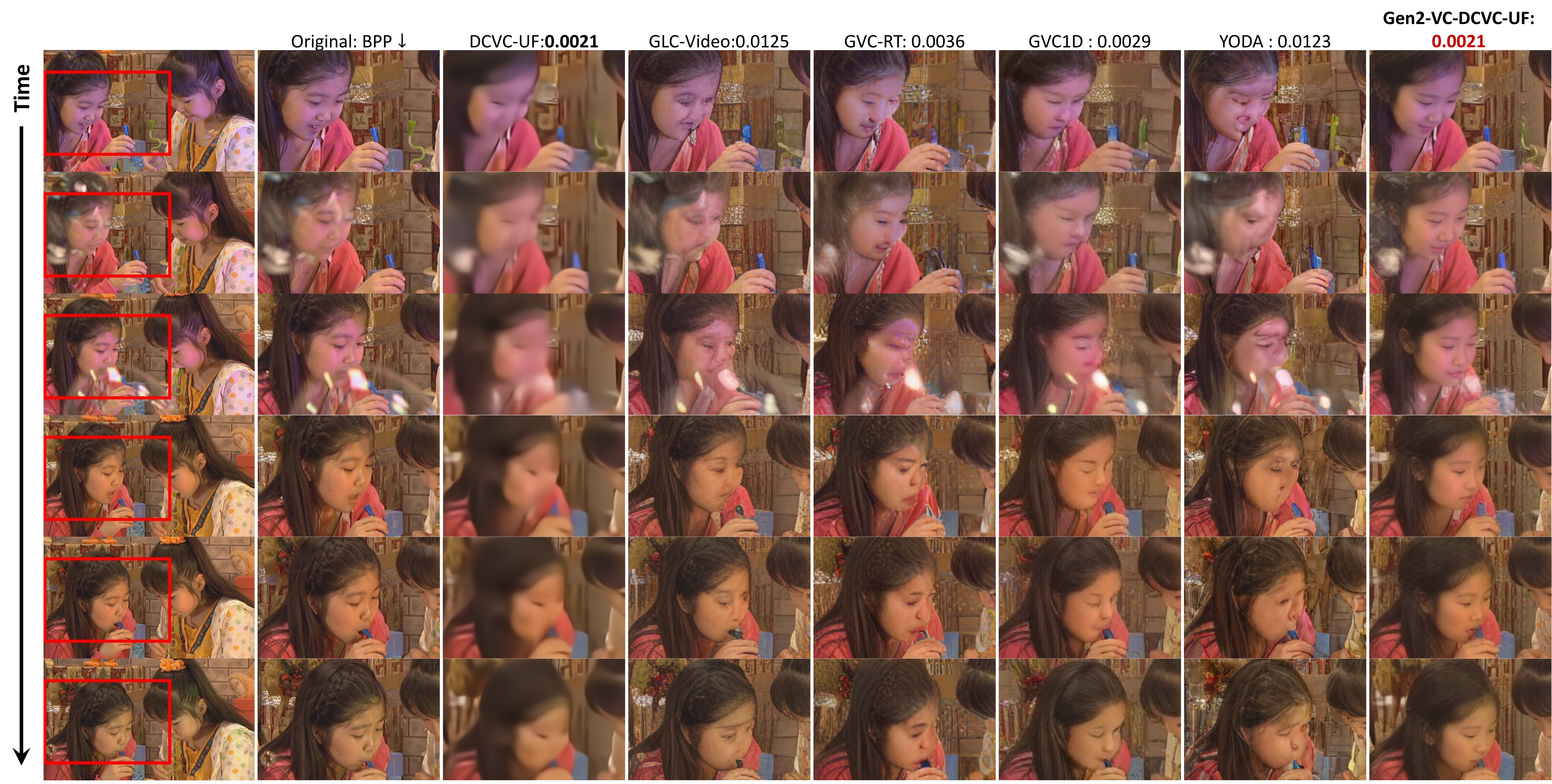}
  \caption{Low-bitrate codec comparison across sampled frames, example~6.}
  \label{fig:app_visual_codec_6}
\end{figure}

\begin{figure}[!htbp]
  \centering
  \includegraphics[width=\linewidth]{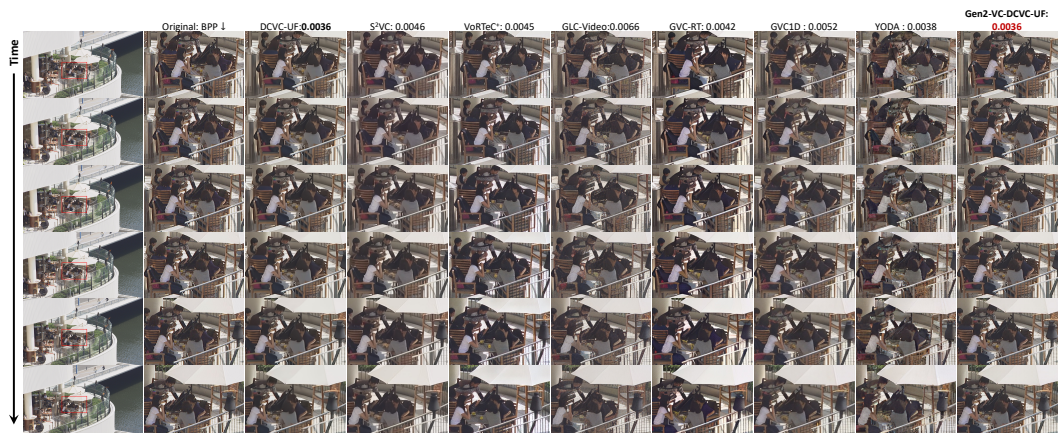}
  \caption{Low-bitrate codec comparison across sampled frames, example~7.}
  \label{fig:app_visual_codec_7}
\end{figure}

Figures~\ref{fig:app_visual_codec_1}--\ref{fig:app_visual_codec_7} compare
Gen2-VC-DCVC-UF with DCVC-UF, GLC-Video, GVC-RT, GVC1D and YODA,
with selected examples also including S$^2$VC and VoRTeC+. 
Gen2-VC recovers finer detail while preserving source structure at the same
bitrate as DCVC-UF and lower bitrates than the generative baselines shown.
Red boxes mark the enlarged regions, BPP is reported above each method,
and sampled frames are ordered from top to bottom.

\subsubsection{Stages and Training-Free Transfer}
\label{app:visual_stages}
\begin{figure}[!htbp]
  \centering
  \includegraphics[width=\linewidth]{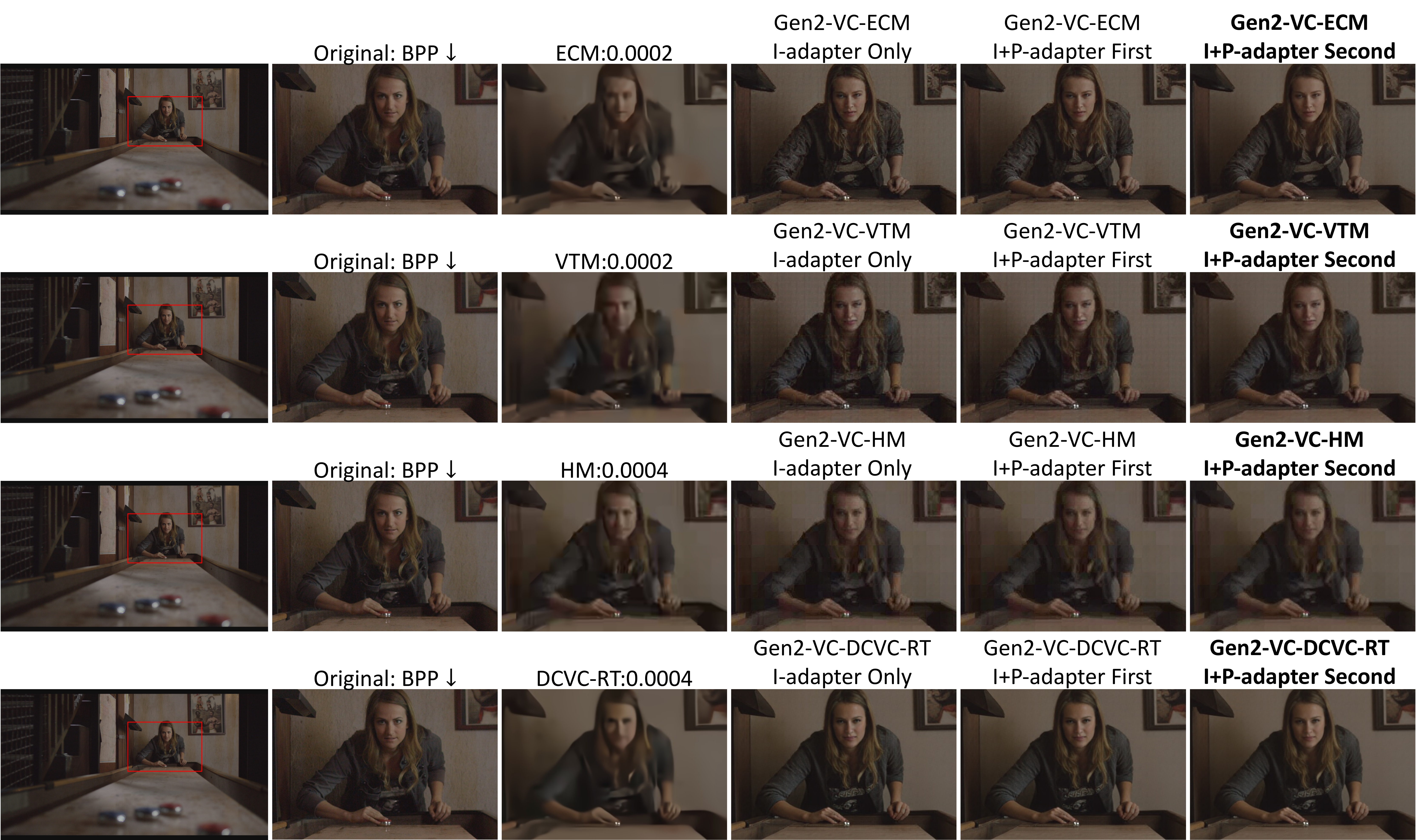}
  \caption{Stage comparisons after training-free transfer to four codecs, example~1.}
  \label{fig:app_visual_stage_1}
\end{figure}

\begin{figure}[!htbp]
  \centering
  \includegraphics[width=\linewidth]{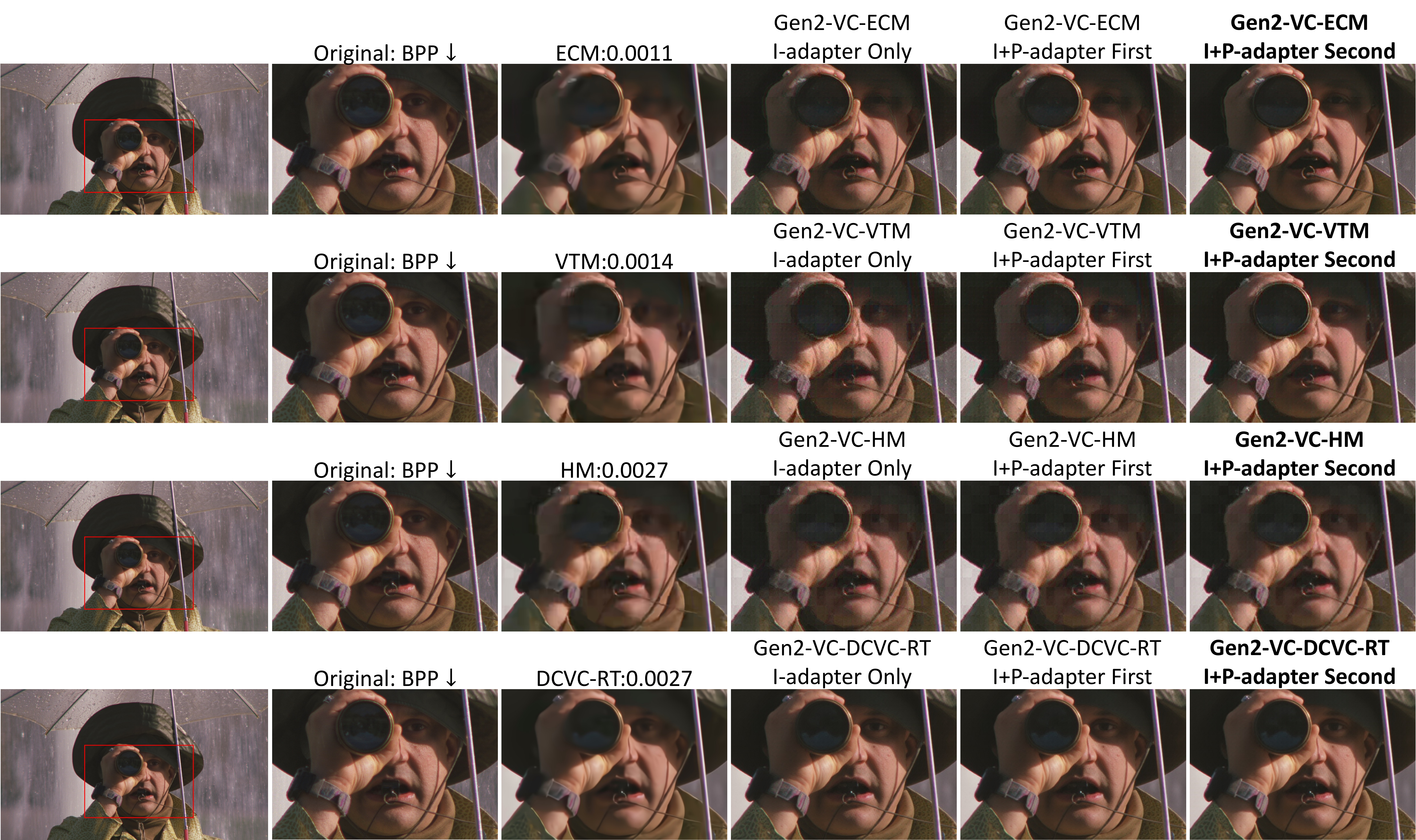}
  \caption{Stage comparisons after training-free transfer to four codecs, example~2.}
  \label{fig:app_visual_stage_2}
\end{figure}

\begin{figure}[!htbp]
  \centering
  \includegraphics[width=\linewidth]{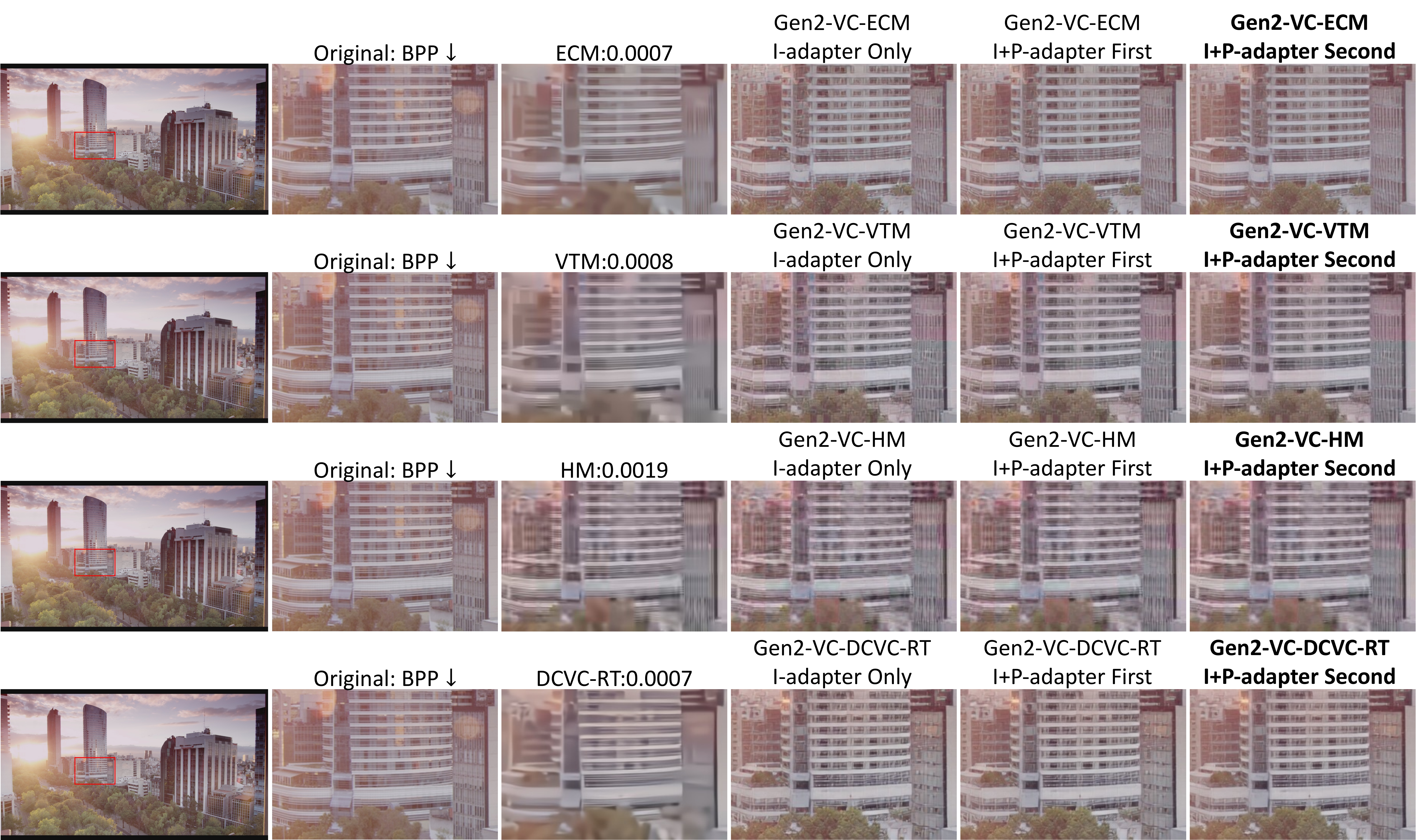}
  \caption{Stage comparisons after training-free transfer to four codecs, example~3.}
  \label{fig:app_visual_stage_3}
\end{figure}

\begin{figure}[!htbp]
  \centering
  \includegraphics[width=\linewidth]{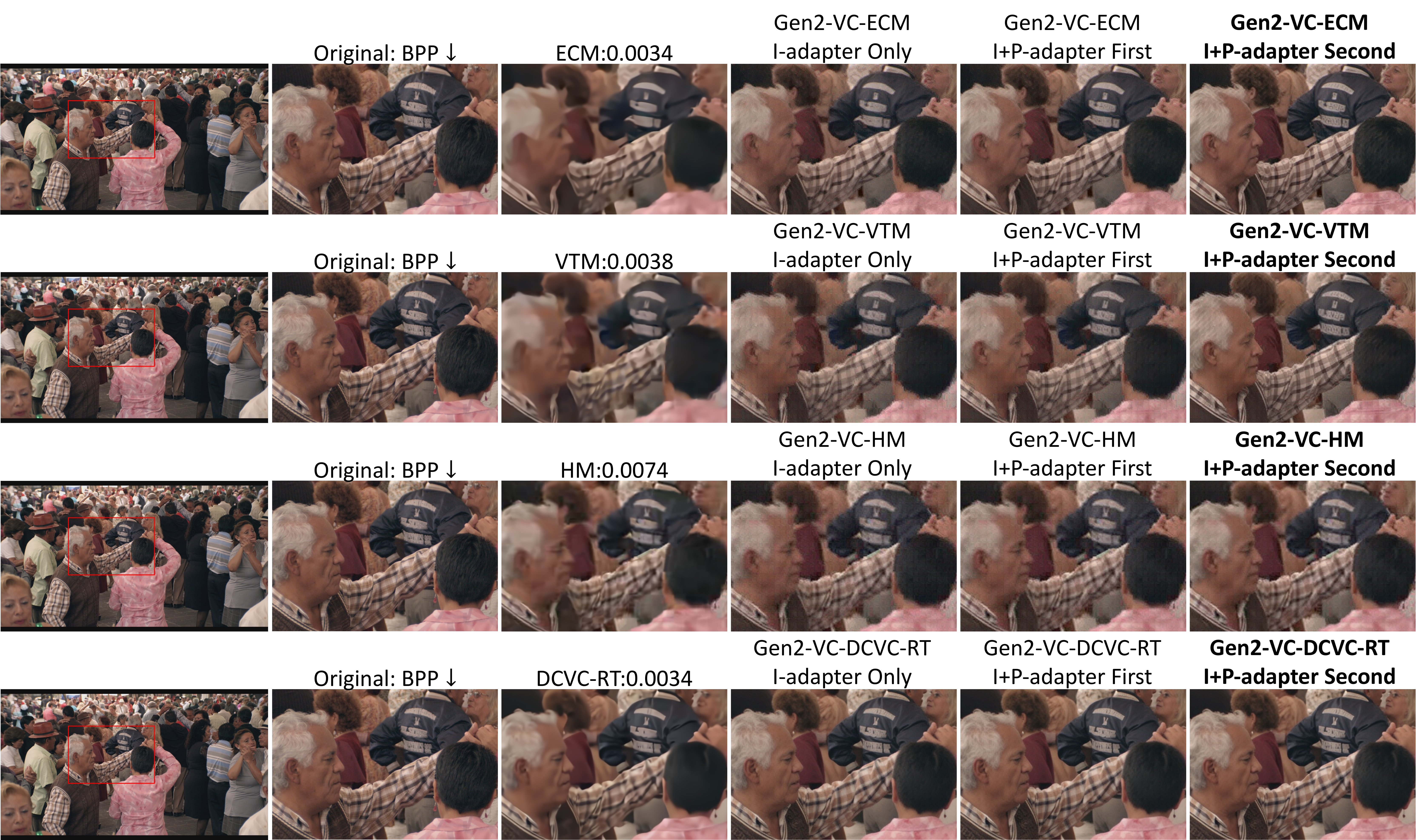}
  \caption{Stage comparisons after training-free transfer to four codecs, example~4.}
  \label{fig:app_visual_stage_4}
\end{figure}

\begin{figure}[!htbp]
  \centering
  \includegraphics[width=\linewidth]{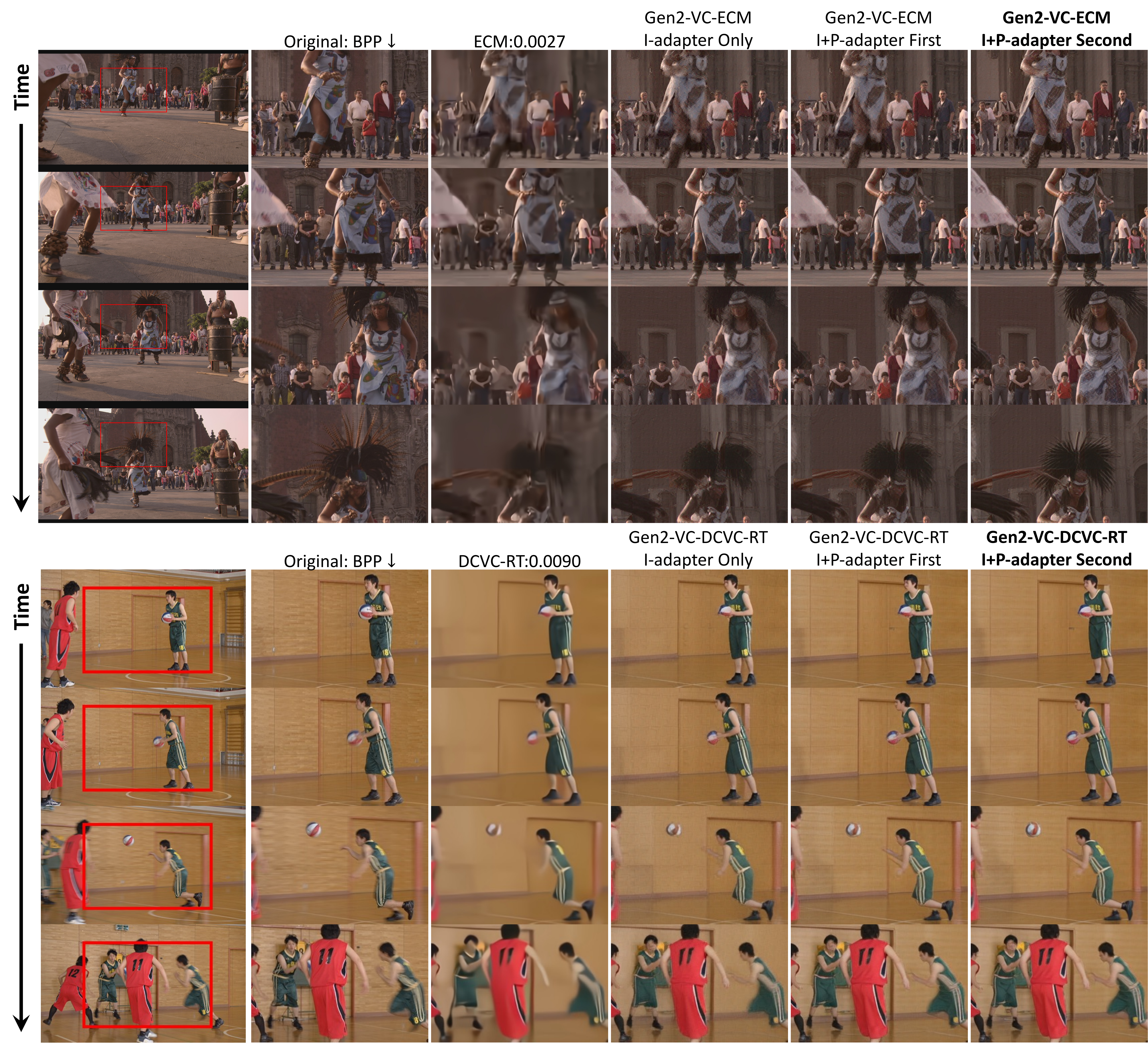}
  \caption{Stage comparisons across sampled frames for Gen2-VC-ECM and Gen2-VC-DCVC-RT.}
  \label{fig:app_visual_stage_5}
\end{figure}

Figures~\ref{fig:app_visual_stage_1}--\ref{fig:app_visual_stage_5} compare
native reconstructions with \textit{I-only} and \textit{I+P} using
\textit{First} or \textit{Second} output selection, following
Figure~\ref{fig:ip_adaptation}.
The adapters transfer from DCVC-UF to ECM, VTM, HM, and DCVC-RT without
retraining, recovering source details across learned and conventional codecs.
Within each codec, all enhancement stages retain the native bitrate.

\subsubsection{Video Restoration Comparisons}
\label{app:visual_restoration}
\begin{figure}[!htbp]
  \centering
  \includegraphics[width=\linewidth]{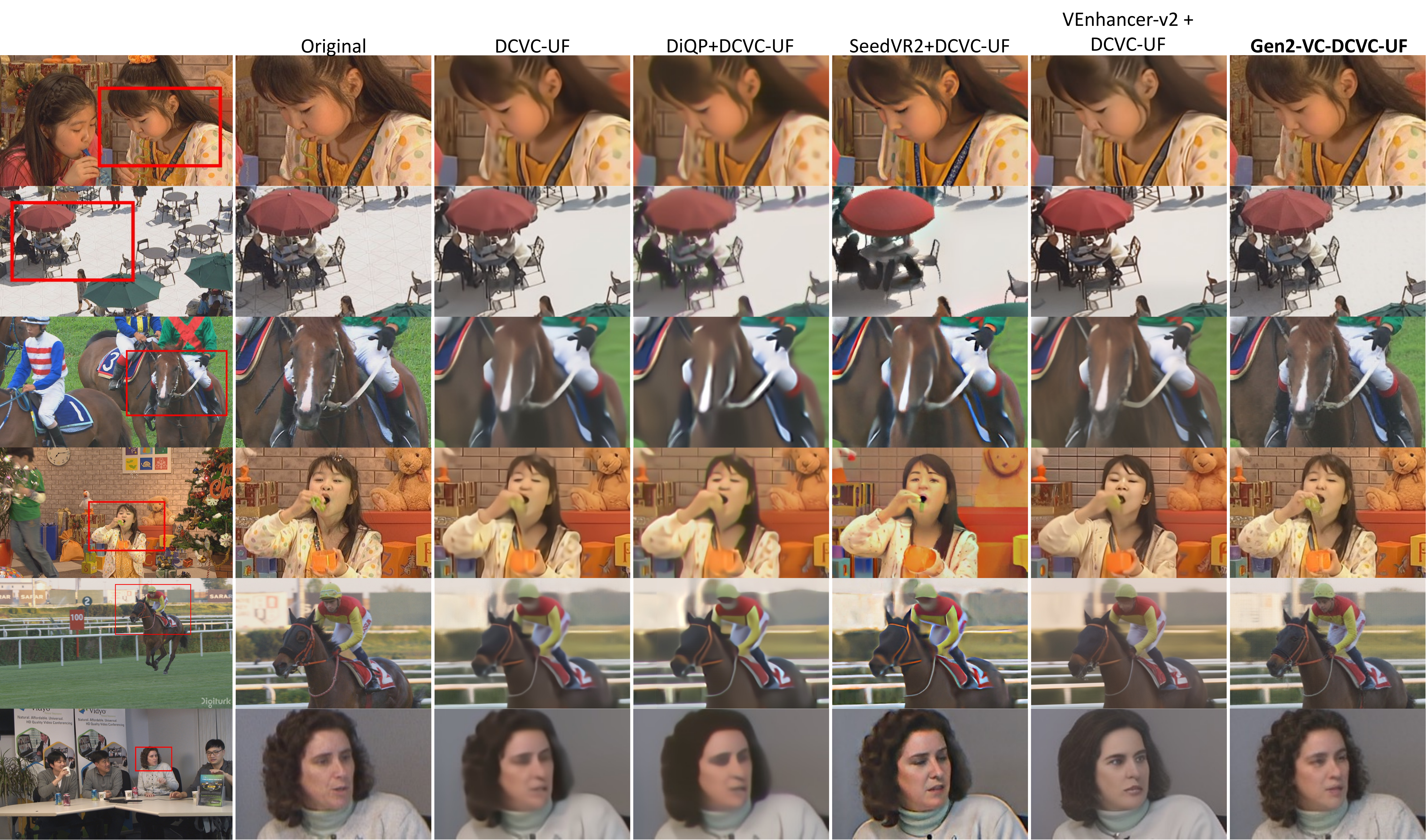}
  \caption{Restoration comparisons on DCVC-UF reconstructions for six selected frames.}
  \label{fig:app_visual_restoration_1}
\end{figure}

\begin{figure}[!htbp]
  \centering
  \includegraphics[width=\linewidth]{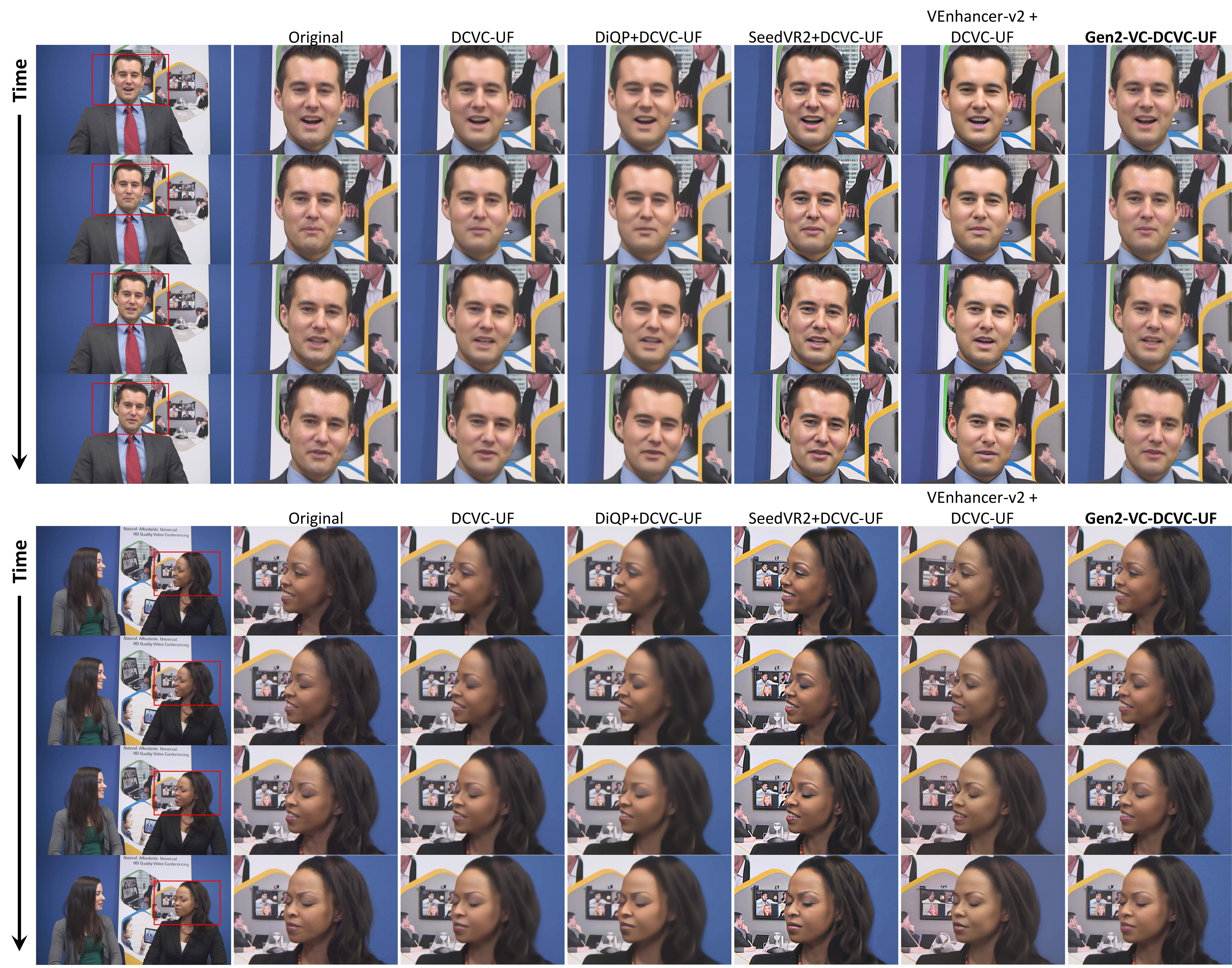}
  \caption{Restoration comparisons on DCVC-UF, with four sampled frames per sequence.}
  \label{fig:app_visual_restoration_2}
\end{figure}

Figures~\ref{fig:app_visual_restoration_1} and~\ref{fig:app_visual_restoration_2}
compare Gen2-VC with DiQP, SeedVR2, and VEnhancer-v2 on the same DCVC-UF
reconstructions.
These examples illustrate Gen2-VC's balance between detail recovery and
source appearance under compression distortions.
Red boxes identify the enlarged regions.
Figure~\ref{fig:app_visual_restoration_2} additionally shows selected frames
in temporal order within each group.

\FloatBarrier

\FloatBarrier

\FloatBarrier

\FloatBarrier

\end{document}